\documentclass[aps,prb,twocolumn,showpacs,superscriptaddress,longbibliography]{revtex4-2}
\usepackage{amsfonts}
\usepackage{amsmath}
\usepackage{graphicx}
\usepackage{bm}
\usepackage{soul}
\usepackage{amssymb}
\usepackage{dcolumn}
\usepackage{color}
\usepackage{multirow}
\usepackage{booktabs}
\usepackage[colorlinks,
linkcolor=blue,
anchorcolor=blue,
citecolor=blue,
urlcolor=blue,
]{hyperref}
\begin{document}

\title{Electro-optic Kerr Effect Induced by Nonlinear Transport in monolayer WTe$_2$}

\author{He-Lin Li}
\affiliation{School of Physical Sciences, University of Chinese Academy of Sciences, Beijng 100049, China}

\author{Zhen-Gang Zhu}
\email{zgzhu@ucas.ac.cn}
\affiliation{School of Electronic, Electrical and Communication Engineering, University of Chinese Academy of Sciences, Beijng 100190, China}

\author{Gang Su}
\email{gsu@ucas.ac.cn}
\affiliation{School of Physical Sciences, University of Chinese Academy of Sciences, Beijng 100049, China}
\affiliation{Kavli Institute for Theoretical Sciences, University of Chinese Academy of Sciences, Beijng 100049, China}
\affiliation{CAS Key Laboratory of Theoretical Physics, Institute of Theoretical Physics, Chinese Academy of Sciences, Beijing 100190, China}

\begin{abstract}
The nonlinear Hall effect (NLHE) can induce optical anisotropy by modifying a material's dielectric tensor, presenting opportunities for novel characterization and device applications. 
While the magneto-optical Kerr effect (MOKE) probes the linear Hall effect (LHE) in magnetic materials, an analogous optical probe for NLHE in non-magnetic, time-reversal symmetric systems remains highly desirable. 
Here, we theoretically propose and investigate an NLHE-induced Electro-optic Kerr Effect (EOKE) as such a probe. Focusing on monolayer (ML) WTe$_2$, a prototypical NLHE material, our analysis considers contributions from Berry curvature dipole (BCD), Drude, injection, and shift mechanisms. We demonstrate that the EOKE signal in WTe$_2$ is predominantly governed by the BCD. Furthermore, the Kerr angle exhibits temporal oscillations at different optical frequencies, suggesting EOKE as a promising route for the time-resolved detection of NLHE and the dynamic investigation of material topology.

\end{abstract}
\maketitle{}

\section{Introduction}
The nonlinear Hall effect (NLHE), characterized by a transverse voltage nonlinearly dependent on the longitudinal driving current, has garnered significant attention for its potential in novel measurement methodologies and device applications \cite{gong2025nonlineartransporttheoryorder,PhysRevLett.115.216806,PhysRevB.92.235447,ma2019observation,PhysRevB.109.085419,du2019disorder,doi:10.1126/sciadv.aay2497,du2021quantum,ho2021hall,PhysRevResearch.2.043081,PhysRevLett.129.186801,he2024third,PhysRevB.110.115123,PhysRevResearch.6.L012059,PhysRevB.109.075419,xiong2025strain}. As a transport phenomenon, the NLHE induces nonlinear electronic responses capable of altering a material's optical properties. Specifically, the current-induced effective fields or charge redistributions associated with NLHE can modify the material's dielectric tensor, particularly its off-diagonal components, thereby inducing optical anisotropy \cite{ZHONG201981,choi2021nonlinearopticalhalleffect,zhang2019optical,doi:10.1126/sciadv.1501524}.

The NLHE is fundamentally distinct from the conventional linear Hall effect (LHE). The LHE is intrinsically linked to Berry curvature (BC) \cite{RevModPhys.82.1959} and requires the breaking of time-reversal symmetry (TRS), typically realized in magnetic materials or under an external magnetic field \cite{RevModPhys.82.1539,nakatsuji2015large,machida2010time,yasuda2016geometric}.
In contrast, the NLHE manifests in non-magnetic materials that preserve TRS. The intrinsic component of the NLHE conductivity tensor is primarily governed by the Berry curvature dipole (BCD) \cite{PhysRevB.97.041101,PhysRevLett.125.046402,PhysRevLett.121.246403}, which emerges from an asymmetric distribution of Berry curvature when spatial inversion symmetry is broken in TRS-invariant systems \cite{PhysRevLett.115.216806,PhysRevB.92.235447}.
Despite these differing symmetry requirements, both effects are rooted in the topological properties of materials and manifested as off-diagonal terms in their respective response tensors. This suggests that detection methods for BC-induced linear Hall conductance might be adaptable for BCD-induced nonlinear Hall conductance, and that nonlinear effects in non-magnetic materials could yield optical phenomena analogous to linear effects in magnetic ones.

In magnetic materials, the LHE is intimately connected to magneto-optical phenomena, exemplified by the magneto-optical Kerr effect (MOKE). MOKE involves the rotation of the polarization angle of linearly polarized light upon reflection from a magnetic material, with the Kerr angle $\theta_k$ being proportional to the linear Hall conductance $\sigma_{xy}$ \cite{Argyres1955,Kerr1875,book}.
While the linear Hall conductivity vanishes in non-magnetic systems with TRS, the Hall response can be expanded to nonlinear orders, resulting in a non-zero nonlinear Hall conductance \cite{sodemann2015quantum,gao2023}.
This raises a compelling prospect: generalizing a MOKE-like effect to non-magnetic materials could provide a powerful tool for detecting the NLHE, thereby opening new avenues to probe the spectral, symmetry, and topological properties of emergent quantum phases \cite{du2021nonlinear}.

Monolayer (ML) WTe$_2$, the first material in which NLHE was experimentally observed, stands as an ideal platform for such investigations \cite{kang2019nonlinear,ma2019observation}. 
Its unique electronic structure, featuring strong spin-orbit coupling, electron-hole compensation, and high carrier mobility, significantly enhances the BCD effect \cite{PhysRevLett.121.266601,kang2019nonlinear,PhysRevLett.115.166601}. Consequently, realizing an MOKE-like effect in WTe$_2$ would be highly representative for advancing the detection of NLHE and exploring material topology.

In this paper, we extend the concept of MOKE from magnetic to non-magnetic systems, proposing a NLHE-induced Electro-optic Kerr Effect (EOKE). 
We theoretically demonstrate a method to detect nonlinear transport using EOKE in ML WTe$_2$. Our analysis, incorporating Drude, BCD, injection, and shift current contributions, reveals that the EOKE signal is predominantly governed by the BCD. Furthermore, by investigating the Kerr effect at different optical frequencies, we find that the total Kerr angle and its BCD-driven component exhibit temporal oscillations. These findings support EOKE as a promising technique for the time-resolved detection of NLHE and the dynamic study of topological properties in materials \cite{PhysRevB.95.094418}.

\section{Electro-optic Kerr Effect with Nonlinear Conductivity}
The MOKE provides a powerful probe of a material's electronic and magnetic properties. 
Conventionally, the Kerr rotation and ellipticity are calculated from the reflection matrix, which is derived using the linear optical conductivity tensor \cite{uselinearkerr,yang2022first}. 
However, in systems with time-reversal symmetry, the linear Hall conductivity vanishes, necessitating the inclusion of nonlinear effects. 
In this work, we extend the formalism to calculate the electro-optic Kerr angle by considering the second-order nonlinear conductivity.
The analysis begins with the wave equation for the electric field $\boldsymbol{E}(\boldsymbol{r},t)$ in a medium, which is derived from Maxwell's equations \cite{Antonov2004ElectronicSA,book}
\begin{equation}
        \nabla(\nabla\cdot\boldsymbol{E}(\boldsymbol{r},t))-\nabla^2\boldsymbol{E}(\boldsymbol{r},t)=-\mu\frac{\partial(\boldsymbol{J}(\boldsymbol{r},t)+\frac{\partial\boldsymbol{D}(\boldsymbol{r},t)}{\partial t})}{\partial t},
        \label{e1}
\end{equation}
where $\mu$ is the magnetic permeability. 
Electric displacement vector $\boldsymbol{D}(\boldsymbol{r},t)=\varepsilon \boldsymbol{E}(\boldsymbol{r},t)$, $\varepsilon$ represents the dielectric tensor.
In this paper, it is expressed as $\varepsilon = \varepsilon_0 \varepsilon_r $.
$\varepsilon_0$ and $\varepsilon_r$ are the permittivity of vacuum and the relative permittivity, respectively. 
The electric field is described as a superposition of plane waves

$
  \boldsymbol{E}(\boldsymbol{r},t)=\sum_{(\boldsymbol{k_i}, \omega_i)= (\pm \boldsymbol{k} , \pm \omega)} \boldsymbol{E} e^{i(\boldsymbol{k}_{i}\cdot \boldsymbol{r}-\omega_{i} t)}, 
$

here, the electric field vector, $\boldsymbol{E} = \sum_{\alpha=x,y,z}E_{\alpha} \boldsymbol{e}_{\alpha}$  represents the complex amplitude of the plane wave. 
The corresponding wave vector is given by $\boldsymbol{k}_{i}=\omega_i \boldsymbol{n}_{i}/c$, where $\boldsymbol{n}_{i}$ is the complex refractive index at frequency $\omega_i$ and $c$ is the speed of light in vacuum.
Here, the subscript i = 1, 2 denotes two distinct modes: i = 1, $n_{1}$ corresponds to $\boldsymbol{k}_{1}$ = $\boldsymbol{k}$ and $\omega_{1}$ = $\omega$;
i = 2, $n_{2}$ corresponds to $\boldsymbol{k}_{2}$ = -$\boldsymbol{k}$ and $\omega_{2}$ = -$\omega$.
The key physical input for our model is the nonlinear current density, which we expand in powers of the electric field strength to the second-order nonlinear term
\begin{align}
  J_{\alpha}(\boldsymbol{r},t) &= \sigma^{\alpha \beta}_{(1)}(\omega_i) E_{\beta} e^{i(\boldsymbol{k}_{i}\cdot \boldsymbol{r}-\omega_{i} t)} \notag\\
  &+ \sum_{\omega_j = \pm \omega}\sigma^{\alpha \gamma \beta }_{(2)}(\omega_i,\omega_j) E_{\gamma}e^{i(\boldsymbol{k}_{j}\cdot \boldsymbol{r}-\omega_{j} t)}E_{\beta}e^{i(\boldsymbol{k}_{i}\cdot \boldsymbol{r}-\omega_{i} t)},
\end{align}
where, the Einstein summation convention over repeated indices ($\beta$, $\gamma$) is used. 

By substituting the relations for $\boldsymbol{J}(\boldsymbol{r},t)$ and $\boldsymbol{D}(\boldsymbol{r},t)$ and applying the plane-wave assumption ($\nabla \to i\boldsymbol{k}$), the wave equation (Eq.~\ref{e1}) is transformed from a differential to an algebraic problem. 
For normal incidence along the z-axis, where the wave vector is $\boldsymbol{k}_i=k_{i,z} \boldsymbol{e}_{z}=(\omega_i n_i / c) \boldsymbol{e}_{z}$, this procedure results in a matrix eigenvalue equation. 
Solving this yields the complex refractive indices $n_{i,\pm}=n_{i,\pm,z}$ for the eigenmodes propagating along the z-axis.  
The explicit expression for these indices is given in Eq.~(\ref{eqnpm}).
The detailed derivation of these indices is provided in the Supplemental Material \cite{SM}.
\begin{widetext}
\begin{eqnarray}
\frac{\omega_i^2}{c^2}n_{i,\pm}^2 - \omega_i^2\mu\varepsilon &=&
\Bigg[-K^{xz}(\omega_i)K^{zx}(\omega_i)
-K^{yz}(\omega_i)K^{zy}(\omega_i)
+K^{xx}(\omega_i)K^{zz}(\omega_i) \notag\\
&+&K^{yy}(\omega_i)K^{zz}(\omega_i)\pm\text{sqrt}
\bigg(\Big[-K^{xz}(\omega_i)K^{zx}(\omega_i)
-K^{yz}(\omega_i)K^{zy}(\omega_i)\notag\\
&+&K^{xx}(\omega_i)K^{zz}(\omega_i)
+K^{yy}(\omega_i)K^{zz}(\omega_i)\Big]^2
+4K^{zz}(\omega_i)\Big[K^{xz}(\omega_i) \notag \\
&\times&K^{yy}(\omega_i)K^{zx}(\omega_i)
-K^{xy}(\omega_i)K^{yz}(\omega_i)K^{zx}(\omega_i)
-K^{xz}(\omega_i) \notag \\
&\times&K^{yx}(\omega_i)K^{zy}(\omega_i)
+K^{xx}(\omega_i)K^{yz}(\omega_i)K^{zy}(\omega_i)
+K^{xy}(\omega_i) \notag \\
&\times&K^{yx}(\omega_i)K^{zz}(\omega_i)
-K^{xx}(\omega_i)K^{yy}(\omega_i)K^{zz}(\omega_i)
 \Big]\bigg)\Bigg]/2K^{zz}(\omega_i)\label{eqnpm}.
\end{eqnarray}
\end{widetext}

To simplify the expression of Eq.~(\ref{eqnpm}), a function $K^{nm}$ is defined as
\begin{align}
  K^{nm}(\omega_i)& \equiv i\omega_i\mu\sigma^{nm}_{(1)}(\omega_i)+ \sum_{\omega_j = \pm \omega}\sum_{\alpha=x,y,z} i(\omega_i+\omega_j)\mu \notag \\
  &\times\sigma^{n\alpha m}_{(2)}(\omega_i+\omega_j)E_\alpha e^{i(\boldsymbol{k}_j\cdot\boldsymbol{r}-\omega_jt)}. \label{Knm}
\end{align}

Eq.~(\ref{Knm}) implies that for the nonlinear conductivity to yield a physically significant contribution to the refractive index, the condition $\omega_i + \omega_j \neq 0$ must be satisfied.
Since $\omega_i$ and $\omega_j$ originate from the same optical field, their frequencies share identical absolute values, i.e., $\omega_i,\omega_j = \pm \omega_j$. 
Crucially, the condition $\omega_i + \omega_j \neq 0$ excludes the case $\omega_i = -\omega_j$. 
Consequently, only the case $\omega_i = \omega_j$ is retained in subsequent calculations herein, corresponding to second-harmonic generation (SHG).

Furthermore, for a two-dimensional (2D) material under the thin-film approximation ($d \ll \lambda$), the analysis of the wave equation yields two optical eigenmodes, which we denote as $\boldsymbol{E}_+$ and $\boldsymbol{E}_-$. 
Their corresponding complex refractive indices are $n_-$ and $n_+$, respectively. 
The explicit expressions for the eigenmodes and their indices are given by Eq.~(\ref{Em}) and Eq.~(\ref{twon}).

\begin{widetext}
\begin{eqnarray}
   \frac{\omega_i^2}{c^2}n_{i,\pm}^2 - \omega_i^2\mu\varepsilon &=& \left(K^{xx}(\omega_i) +K^{yy}(\omega_i)\right)/2 \notag{}\\
   &\pm& \left( K^{xx}(\omega_i)^2 +4K^{xy}(\omega_i)K^{yx}(\omega_i) \right. - \left. 2K^{xx}(\omega_i)K^{yy}(\omega_i)+K^{yy}(\omega_i)^2 \right)^{\frac{1}{2}}/2.
\label{twon}
\end{eqnarray}
\end{widetext}

\begin{widetext}
\begin{align}
  &\boldsymbol{E}_{\mp} = |\boldsymbol{E}|\notag \\& \times\left(\boldsymbol{e}_y  -\left( \frac{\left(-K^{xx}(\omega_i) + K^{yy}(\omega_i) \mp \sqrt{K^{xx}(\omega_i)^2 + 4 K^{xy}(\omega_i) K^{yx}(\omega_i) - 2 K^{xx}(\omega_i) K^{yy}(\omega_i) + K^{yy}(\omega_i)^2}\right)}{2 K^{yx}(\omega_i)} \right) \boldsymbol{e}_x \right). 
  \label{Em}
  \end{align}
\end{widetext}
Throughout this work, the electric field intensity E refers to the amplitude $|\boldsymbol{E}|$ from Eq.~\ref{Em}.

The incident light can always be decomposed into these two modes by rotating the polarization direction of the incident light.
For light at normal incidence from a vacuum ($n_0 = 1$) onto the material, the reflection coefficients for these two eigenmodes, $r_{i,+}$ and $r_{i,-}$, are given by the standard Fresnel formula
\begin{align}
  r_{i,\pm}=\frac{n_0-n_{i,\pm}}{n_{i,\pm}+n_0}.
\end{align}

An incident linearly polarized light can be decomposed into these two eigenmodes. 
Upon reflection, the difference between the complex reflection coefficients $r_{i,+}$ and $r_{i,-}$ introduces a phase shift and an amplitude change between the two components. 
This rotates the polarization of the reflected light, an effect known as the Kerr effect.
This polarization change is quantified by the Kerr angle, $\theta_k$. 
It is determined by the ratio of the reflection coefficients of the two eigenmodes. 
For normal incidence, it is defined as
\begin{align}
  \frac{\boldsymbol{E}_{+,out}}{\boldsymbol{E}_{-,out}}&=\frac{n_{i,-}-n_0}{n_{i,-}+n_0}\frac{n_{i,+}+n_0}{n_{i,+}-n_0} \notag\\
   &=\frac{|\boldsymbol{E}_{+,out}|}{|\boldsymbol{E}_{-,out}|}e^{i(\phi_{+}-\phi_{-})} \notag\\
   &=\frac{|\boldsymbol{E}_{+,out}|}{|\boldsymbol{E}_{-,out}|}e^{\frac{1}{2}i(\theta_k)},
\end{align}
where ${\boldsymbol{E}_{+,out}}$, $\boldsymbol{E}_{-,out}$ is a two different mode of reflected light with reflective index $n_{i,-}$ and $n_{i,+}$.

Because the expression of the electric field is represented as the sum of two frequencies $\pm \omega$, the Kerr angle can be expressed as
\begin{eqnarray}
    \theta_k &=&\frac{1}{4}\left[\text{Arg}\left(\frac{n_--n_0}{n_-+n_0} \frac{n_++n_0}{n_+-n_0}\right)|_{\omega<0} \right. \notag \\
    &+& \left. \text{Arg}\left(\frac{n_--n_0}{n_-+n_0} \frac{n_++n_0}{n_+-n_0}\right)|_{\omega>0}\right].
\end{eqnarray}

\section{nonlinear conductivity}

In this chapter, we calculate the conductivity using quantum kinetics \cite{matsyshyn2019nonlinear,liu2024photogalvanic}. 
The light-induced current density, $\langle j^\eta \rangle$, can be expanded up to the second order in the electric field

\begin{align}
  \langle j^\eta \rangle     &=\langle j^\eta \rangle^{(1)} + \langle j^\eta \rangle^{(2)}, \\
  \langle j^\eta\rangle^{(1)}&=\sum_{\omega_i = \pm \omega} \sigma_{(1)}^{\eta\alpha}\left(\omega_i\right)E_\alpha\left(\omega_i\right)e^{-i\omega_i t}, \\
  \langle j^\eta\rangle^{(2)}&=\sum_{\omega_i, \omega_j = \pm \omega}\sigma_{(2)}^{\eta\alpha\beta}\left(\omega_i,\omega_j\right)E_\alpha\left(\omega_i\right)\notag\\
                             &\times E_\beta\left(\omega_j\right)e^{-i(\omega_i+\omega_j)t}.
\end{align}
Here, the Einstein summation convention over repeated indices ($\alpha$, $\beta$) is used. 
Where $\alpha$, $\beta$, $\eta$ refer to the Cartesian component.
$E_{\alpha}$ and $E_{\beta}$ are the amplitudes of the electric field components with corresponding frequencies $\omega_{i}$ and $\omega_{j}$ (where $\omega_{i} = \pm \omega$, $\omega_{j} = \pm \omega$). 

The linear conductivity tensor $\sigma_{(1)}^{\eta\alpha}$ is given by
\begin{align}
  \sigma_{(1)}^{\eta\alpha} = \frac{e^2}{\hbar}\int\frac{d^3k}{(2\pi)^3}\sum_nf_n\Omega_n^{\alpha\eta}.  
\end{align}
In this expression, $f_n$ is the Fermi-Dirac distribution function for the $n$-th band.
The term $\Omega_n^{\alpha\eta}$ is the Berry Curvature, defined as
\begin{equation}
  \Omega_{n}^{\alpha\eta}(\boldsymbol{k})\equiv i\left(\langle\partial_{\alpha}u_{n\boldsymbol{k}}|\partial_{\eta}u_{n\boldsymbol{k}}\rangle-\langle\partial_{\eta}u_{n\boldsymbol{k}}|\partial_{\alpha}u_{n\boldsymbol{k}}\rangle\right),
\end{equation}
where $|u_{n\boldsymbol{k}}\rangle$ is the periodic part of the Bloch wavefunction.

The second-order conductivity, $\sigma_{(2)}^{\eta\alpha\beta}\left(\omega_i,\omega_j\right)$, can be decomposed into four main contributions: the BCD, Drude, shift, and injection currents \cite{liu2024photogalvanic, Sipe2000}. 
By incorporating a finite relaxation time, $\tau=1/\Gamma$ (where $\Gamma$ is the relaxation rate), these terms are given by:

BCD term
\begin{align}
  \sigma_{(2)\mathrm{BCD}}^{\eta\alpha\beta}\left(\omega_i,\omega_j\right)  &= \frac{e^3}{\hbar^3} \int \frac{d^3k}{(2\pi)^3} \sum_{nm} d^{\omega_i} d_{nm}^{\omega_i+\omega_j} \notag\\
  &\times  \varepsilon_{mn}\xi_{mn}^\eta\xi_{nm}^\beta\partial_{k_\alpha}f_{mn}(\boldsymbol{k}), \label{eqsigmaBCD}
\end{align}

Drude term 
\begin{align}
  \sigma_{(2)\mathrm{Drude}}^{\eta\alpha\beta}\left(\omega_i,\omega_j\right) &= \frac{e^3}{\hbar^3}\int\frac{d^3k}{(2\pi)^3} \sum_nd^{\omega_i} d^{\omega_i+\omega_j} \notag \\
  &\times  \left(\partial_{k_\eta}\varepsilon_n\right)\partial_{k_\alpha}\partial_{k_\beta}f_n(\boldsymbol{k}), \label{eqsigmaDrude}
\end{align}

Injection term 
\begin{align}
  \sigma_{(2)\mathrm{Injection}}^{\eta\alpha\beta}\left(\omega_i,\omega_j\right)  &= -\frac{e^3}{2\hbar^3}\int\frac{d^3k}{(2\pi)^3} \sum_{nm}\hbar d^{\omega_i+\omega_j} \left(d_{nm}^{\omega_i} \right. \notag\\
  &+\left. d_{mn}^{\omega_j}\right)f_{nm}(\boldsymbol{k})\Delta_{nm}^\eta\xi_{mn}^\beta\xi_{nm}^\alpha, \label{eqsigmaInjection}
\end{align}

Shift term
\begin{align}
&\sigma_{(2)\mathrm{Shift}}^{\eta\alpha\beta}  \left(\omega_i,\omega_j\right) \notag\\
&= \frac{e^3}{\hbar^3}\int\frac{d^3k}{(2\pi)^3}\sum_{nm} \Biggl\{d_{nm}^{\omega_i+\omega_j}\varepsilon_{mn}\xi_{mn}^\eta\xi_{nm}^\alpha \partial_{k_\beta}\left[d_{nm}^{\omega_i}f_{mn}(\boldsymbol{k})\right]  \notag \\
  &+  d_{nm}^{\omega_i}d_{nm}^{\omega_i+\omega_j}\varepsilon_{mn}  f_{mn}\xi_{mn}^\eta\xi_{nm;\beta}^\alpha - id_{nm}^{\omega_i+\omega_j}\varepsilon_{mn}\xi_{mn}^\eta \notag \\
  &\times \sum_{l\neq n,m}\left(d_{lm}^{\omega_i}\xi_{lm}^\alpha\xi_{nl}^\beta f_{ml}-d_{nl}^{\omega_i}\xi_{nl}^\alpha\xi_{lm}^\beta f_{ln}\right)\Biggr\}.\label{eqsigmaShift}
\end{align}

The terms in the preceding equations are defined as follows. 
The energy of the $n$-th band is $\varepsilon_{n}$, and the corresponding Fermi-Dirac distribution is $f_{n}$.
Their differences between bands $n$ and $m$ are written as $\varepsilon_{nm} \equiv \varepsilon_{n} - \varepsilon_{m}$ and $f_{nm} \equiv f_{n} - f_{m}$.
The velocity difference, $\Delta_{nm}^\alpha$,  is related to the energy gradient by $\Delta_{nm}^\alpha\equiv\upsilon_{nn}^\alpha-\upsilon_{mm}^\alpha=\frac{1}{\hbar}\frac{\partial\varepsilon_{nm}}{\partial k_\alpha}$.
The frequency and relaxation-dependent denominators are given by
$d^\omega\equiv\frac{1}{\omega+i\Gamma}$ and $d_{nm}^\omega\equiv\frac{1}{\omega-\varepsilon_{nm}/\hbar+i\Gamma}$. 
Finally, the geometric quantities include the non-Abelian Berry connection, defined as $\boldsymbol{\xi}_{nm}(\boldsymbol{k}) \equiv i\langle u_{n\boldsymbol{k}}|\partial_{\boldsymbol{k}}u_{m\boldsymbol{k}}\rangle$ \cite{matsyshyn2019nonlinear,liu2024photogalvanic},
and its generalized derivative, $\boldsymbol{\xi}_{nm}(\boldsymbol{k})$ is defined as $\xi_{nm;\alpha}^\beta \equiv \partial_{k_\alpha}\xi_{nm}^\beta - i(\xi_{nn}^\alpha-\xi_{mm}^\alpha)\xi_{nm}^\beta$ \cite{Sipe2000}.

\section{RESULTS AND DISCUSSION}
\subsection{{Kerr angle in WTe$_2$ monolayer}}
%%%%%%%%%%%%%%%%%%%%%%%%%%%%%%%%%%%%%%%%%%%%%%%%%%%%%%%%%%%%%%%%%%%%%%%%%%%%%%%%%%%%%%%%%%%%%%%%%%%%%%%%%%%%%
\begin{figure}[tb]
  \centering
    \includegraphics[width=1\columnwidth]{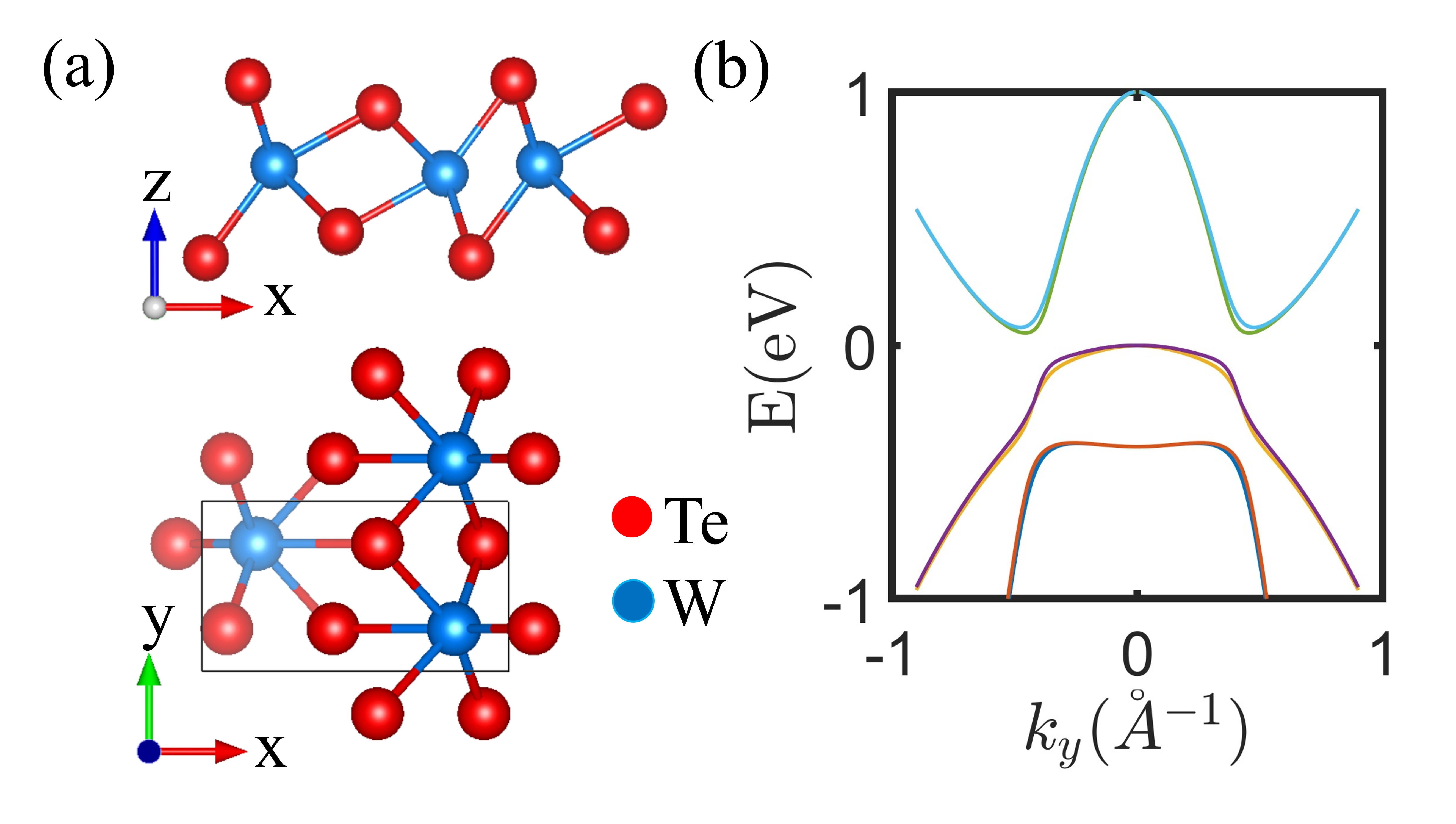}
    \caption{(a) Schematic of a single layer of WTe$_2$. 
    (b) Band Structure calculated along the $k_y$ direction at $k_x = 0$ with parameters $\delta_{3,x}$ = $\delta_{3,z}$ $\thickapprox$ 0, and $\delta_{1,x}$ $\thickapprox$ 0, $\delta_{1,z}$ = 0.025 eV. }
  \label{fig_WTe2_band}
\end{figure}
%%%%%%%%%%%%%%%%%%%%%%%%%%%%%%%%%%%%%%%%%%%%%%%%%%%%%%%%%%%%%%%%%%%%%%%%%%%%%%%%%%%%%%%%%%%%%%%%%%%%%%%%%%%%%

ML WTe$_2$ is a prominent two-dimensional topological insulator whose properties can be controlled by an out-of-plane electric field, making it an ideal platform for investigating novel quantum phenomena \cite{WTe2TI,WTe2BCD}. 
WTe$_2$ exists in two closely related structures, the 1T$'$ and T$_d$ phases. In this work, we focus on the T$_d$ phase, depicted in Fig.~\ref{fig_WTe2_band}(a).

The six-band model for monolayer T$_d$-WTe$_2$, introduced by Shi et al. (2019) \cite{WTE2H}, accurately describes the distribution of geometric quantities across the entire commutation space. Thus, this model is adopted in our study.
The $\boldsymbol{k} \cdot \boldsymbol{p}$ Hamiltonian reads \cite{liu2024photogalvanic}
\begin{equation}
    \left.H_0(\boldsymbol{k})=\left(
    \begin{array}{cccccc}
    \epsilon_1 & \upsilon_1^+ & 0 & 0 & 0 & 0 \\
    -\upsilon_1^- & \epsilon_2 & \upsilon_3^+ & 0 & 0 & 0 \\
    0 & -\upsilon_3^- & \epsilon_3 & 0 & 0 & 0 \\ 0 & 0 & 0 & \epsilon_1 & \upsilon_1^- & 0 \\ 0 & 0 & 0 & -\upsilon_1^+ & \epsilon_2 & \upsilon_3^- \\ 0 & 0 & 0 & 0 & -\upsilon_3^+ & \epsilon_3
    \end{array}
    \right.
    \right).
\end{equation}

Among as $\epsilon_i=c_{i,0}+c_{i,x}k_x^2+c_{i,y}k_y^2\mathrm{~and~}v_i^\pm=\pm v_{i,x}k_x+iv_{i,y}k_y$ for $i$-th orbital. The parameters are taken from Table III in Ref.~\cite{WTE2H}. The effect of an external vertical electric field is represented through the term $H_{1}(\boldsymbol{k})$ as 
\begin{equation}
  H_1(\boldsymbol{k})=
  \begin{pmatrix}
  0 & i\delta_{1,z} & 0 & 0 & i\delta_{1,x} &  0  \\
  -i\delta_{1,z} & 0 & i\delta_{3,z} & -i\delta_{1,x} & 0 &  i\delta_{3,x}  \\
  0 & -i\delta_{3,z} & 0 & 0 & -i\delta_{3,x} &  0  \\
  0 & i\delta_{1,x}  & 0 & 0 & -i\delta_{1,z} &  0  \\
  -i\delta_{1,x}     & 0 & i\delta_{3,x}  &  i\delta_{1,z}  & 0 & -i\delta_{3,z} \\
  0  &  -i\delta_{3,x}  &  0  &  0  &  i\delta_{3,z}  &  0
  \end{pmatrix},
\end{equation}
where $\delta_{i,x}$ and $\delta_{i,z}$ are respectively related to the Rashba and Ising spin-orbit couplings induced by the applied perpendicular electric field, and they are both $\mathbf{k}$ independent. The total Hamiltonian is then 
\begin{equation}
  H(\boldsymbol{k})=H_0(\boldsymbol{k})+H_1(\boldsymbol{k}).
\label{e13}
\end{equation}
Fig.~\ref{fig_WTe2_band} (b) shows the band structure diagram corresponding to Eq. (\ref{e13}).

\subsection{Nonlinear Electro-optic Kerr Effect}
Since WTe$_2$ is a non-magnetic material, we neglect magnetization effects in this study and assume the permeability to be that of a vacuum, $\mu = \mu_0$. 
The relative permittivity is set to $\varepsilon_r = 3.3$, following Ref.~\cite{KUMAR20124627}.

\begin{figure}[tb]
\centering
\includegraphics[scale=0.75,angle=0,width=8.35cm,height=8cm]{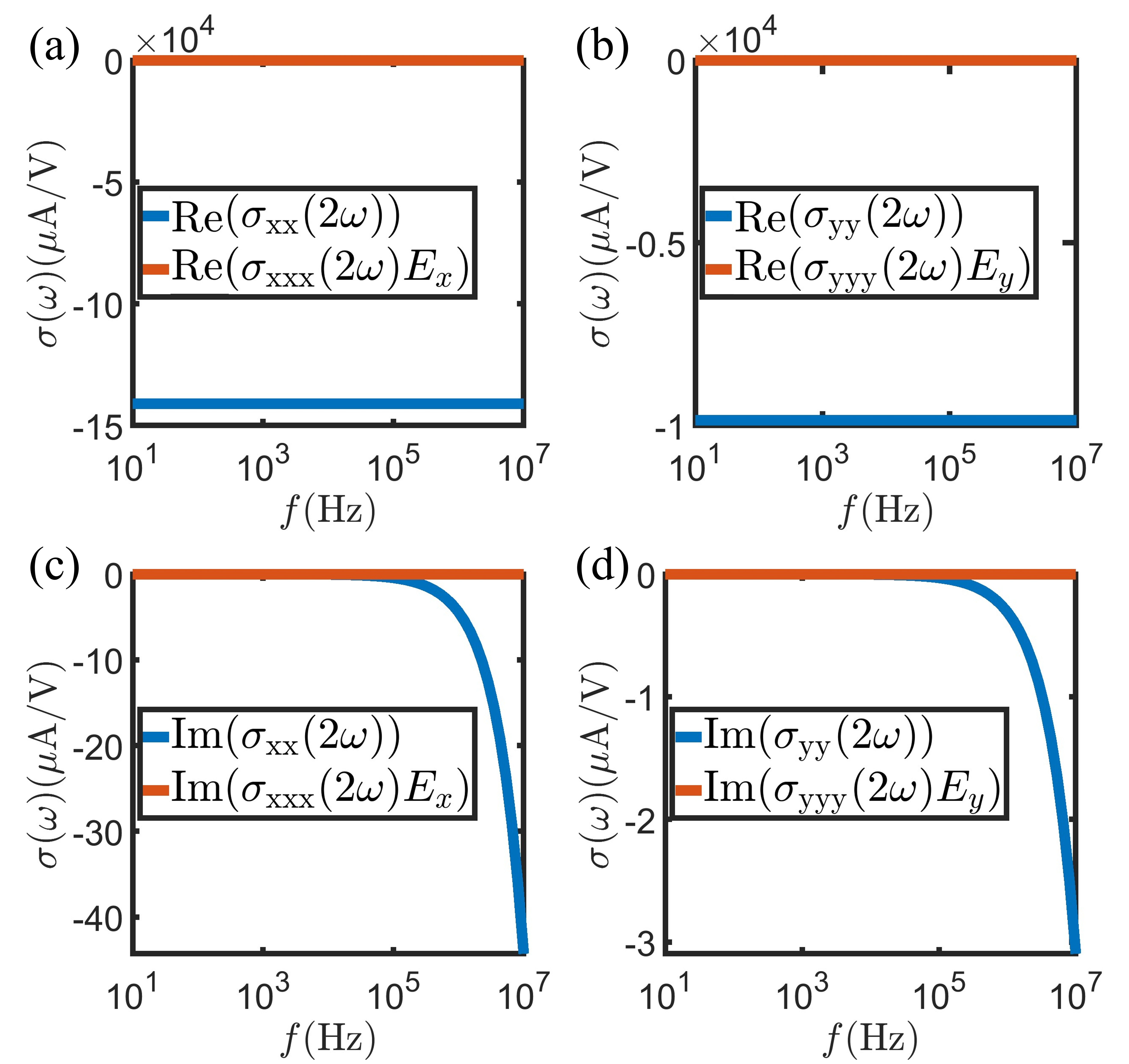}\\
\caption{
  Linear and nonlinear longitudinal conductivity as a function of frequency, calculated with $\tau= 5$ ps and $E_f = 0.12$ eV.
(a) Real parts of the linear ($\sigma_{xx}$) and nonlinear ($\sigma_{xxx}$) conductivities.
(b) Real parts of the linear ($\sigma_{yy}$) and nonlinear ($\sigma_{yyy}$) conductivities.
(c) Imaginary parts of $\sigma_{xx}$ and $\sigma_{xxx}$.
(d) Imaginary parts of $\sigma_{yy}$ and $\sigma_{yyy}$.
The nonlinear components ($\sigma_{xxx}$, $\sigma_{yyy}$) were calculated with an electric field strength of 1 V/nm ($E_x$, $E_y$ = 1 V/nm).
  }
\label{fig_sigma_xx}
\end{figure}

Our calculations are performed under the conditions 
$\sigma^{nm}_{(1)}(\omega_i) \gg \sigma^{n\alpha m}_{(2)}(\omega_i+\omega_j)$ 
and $|E_{\alpha}| \ll 1\,\mathrm{V/nm}$ (see Fig.~\ref{fig_sigma_xx}).
This necessitates a specific criterion for the Kerr coefficient $K^{nm}(\omega_i)$: 
contributions from the second-order conductivity, $\sigma^{n\alpha m}_{(2)}(\omega_i+\omega_j)$, are included only when the first-order term $\sigma^{nm}_{(1)}(\omega_i)$ vanishes.
%
%%%%%%%%%%%%%%%%%%%%%%%%%%%%%%%%%%%%%%%%%%%%%%%%%%%%%%%%%%%%%%%%%%%%%%%%%%%%%%%%%%%%%%%%%%%%%%%%%%%%%%%%%%%%%%%%
%%%%%%%%%%%%%%%%%%%%%%%%%%%%%%%%%%%%%%%%%%%%%%%%%%%%%%%%%%%%%%%%%%%%%%%%%%%%%%%%%%%%%%%%%%%%%%%%%%%%%%%%%%%%%%%%
%
%%%%%%%%%%%%%%%%%%%%%%%%%%%%%%%%%%%%%%%%%%%%%%%%%%%%%%%%%%%%%%%%%%%%%%%%%%%%%%%%%%%%%%%%%%%%%%%%%%%%%%%%%%%%%%%%
\begin{figure*}[tb]
\centering
    \includegraphics[scale=0.4,angle=0,width=16.7cm]{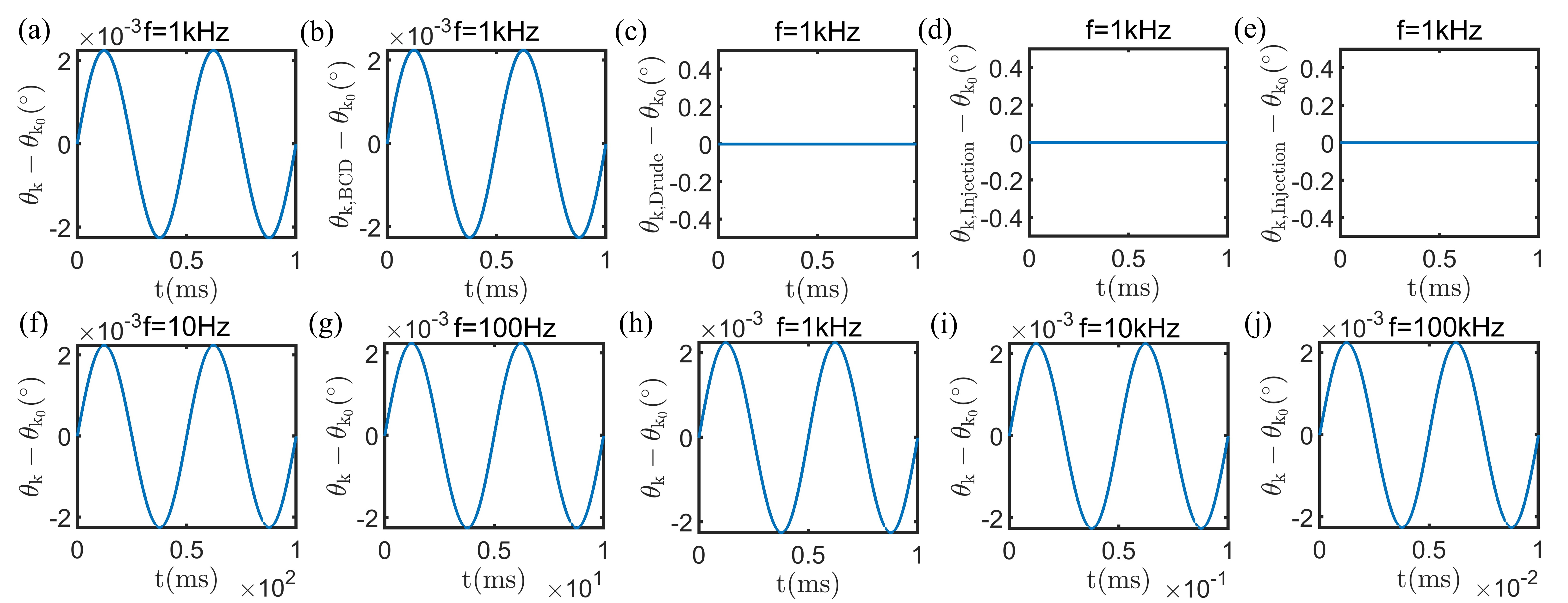}\\
    \caption{
      Temporal variation of the electro-optic Kerr angle with parameters $\tau = 5~\mathrm{ps}$, $E_f = 0.12~\mathrm{eV}$, and $E = 10^{-1}~\mathrm{V/nm}$. 
        (a-e) Total Kerr angle and its time-dependent components at a light frequency $f = 1~\mathrm{kHz}$.
        (f-j) Total Kerr angle at different frequencies: $f = 10~\mathrm{Hz}$, $10^2~\mathrm{Hz}$, $10^3~\mathrm{Hz}$, $10^4~\mathrm{Hz}$, $10^5~\mathrm{Hz}$.
        Here, $\theta_{k_0}$ denotes the Kerr rotation angle arising from the in-plane conductivity anisotropy ($\sigma_{xx} \neq \sigma_{yy}$) within the linear response regime.
        }
\label{fig_Kerr_t}
\end{figure*}
%%%%%%%%%%%%%%%%%%%%%%%%%%%%%%%%%%%%%%%%%%%%%%%%%%%%%%%%%%%%%%%%%%%%%%%%%%%%%%%%%%%%%%%%%%%%%%%%%%%%%%%%%%%%%%%%

%
%%%%%%%%%%%%%%%%%%%%%%%%%%%%%%%%%%%%%%%%%%%%%%%%%%%%%%%%%%%%%%%%%%%%%%%%%%%%%%%%%%%%%%%%%%%%%%%%%%%%%%%%%%%%%%%%
\begin{figure*}[tb]
\centering
    \includegraphics[scale=0.4,angle=0,width=16.7cm]{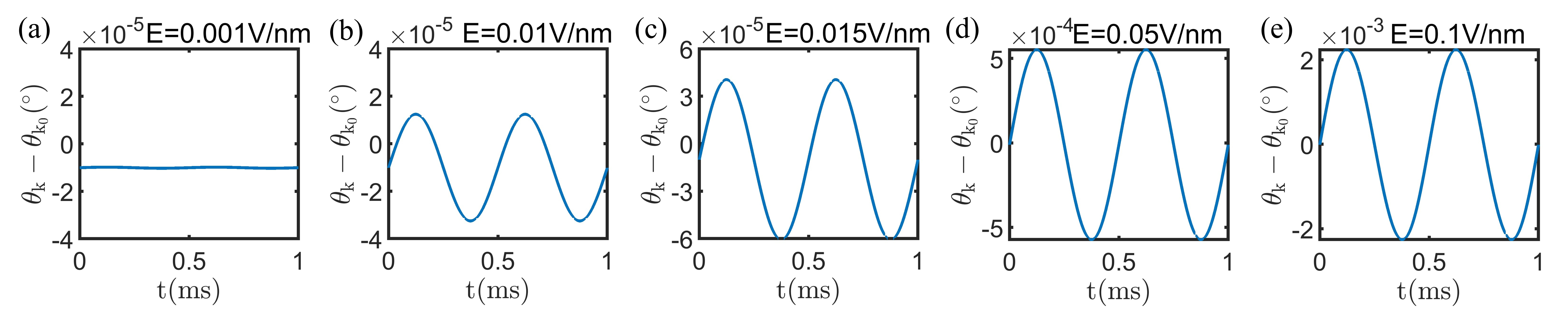}
    \caption{
      Temporal variation of the total electro-optic Kerr angle under different electric field strengths, with fixed parameters $\tau = 5~\mathrm{ps}$ and $E_f = 0.12~\mathrm{eV}$ and $f = 1~\mathrm{kHz}$. 
        (a-e) Results for electric field strengths $E = 1*10^{-3}~\mathrm{V/nm}$, $1*10^{-2}~\mathrm{V/nm}$, $1.5*10^{-2}~\mathrm{V/nm}$, $0.5*10^{-1}~\mathrm{V/nm}$, and $10^{-1}~\mathrm{V/nm}$.
        Here, $\theta_{k_0}$ denotes the Kerr rotation angle arising from the in-plane conductivity anisotropy ($\sigma_{xx} \neq \sigma_{yy}$) within the linear response regime.
        }
\label{fig_Kerr_E}
\end{figure*}
%%%%%%%%%%%%%%%%%%%%%%%%%%%%%%%%%%%%%%%%%%%%%%%%%%%%%%%%%%%%%%%%%%%%%%%%%%%%%%%%%%%%%%%%%%%%%%%%%%%%%%%%%%%%%%%%
%
Applying the formalism described above, we now present the calculated nonlinear EOKE (NEOKE) for ML WTe$_2$.
Fig.~\ref{fig_Kerr_t} displays the calculated using a relaxation time of $\tau$ = 5 ps and an electric field amplitude of $|E|$ = $10^{-1}$ V/nm.
The total Kerr angle, shown in Fig.~\ref{fig_Kerr_t}(a), incorporates four distinct mechanisms: BCD, Drude, Injection, and Shift currents, such that the total second-order conductivity is $\sigma_{(2)} = \sigma_{(2)\text{BCD}}+\sigma_{(2)\text{Drude}}+\sigma_{(2)\text{Injection}}+\sigma_{(2)\text{Shift}}$.
To clarify the origin of the total signal, we also analyze the contribution from each mechanism individually. 
For instance, the BCD Kerr angle refers to the angle induced solely by the BCD effect ($\sigma_{(2)} = \sigma_{(2)\text{BCD}} $), with other contributions omitted.
These individual contributions are detailed in Figs.~\ref{fig_Kerr_t}(b-e).

\begin{figure}[tb]
\centering
\includegraphics[scale=0.75,angle=0,width=8.35cm,height=4cm]{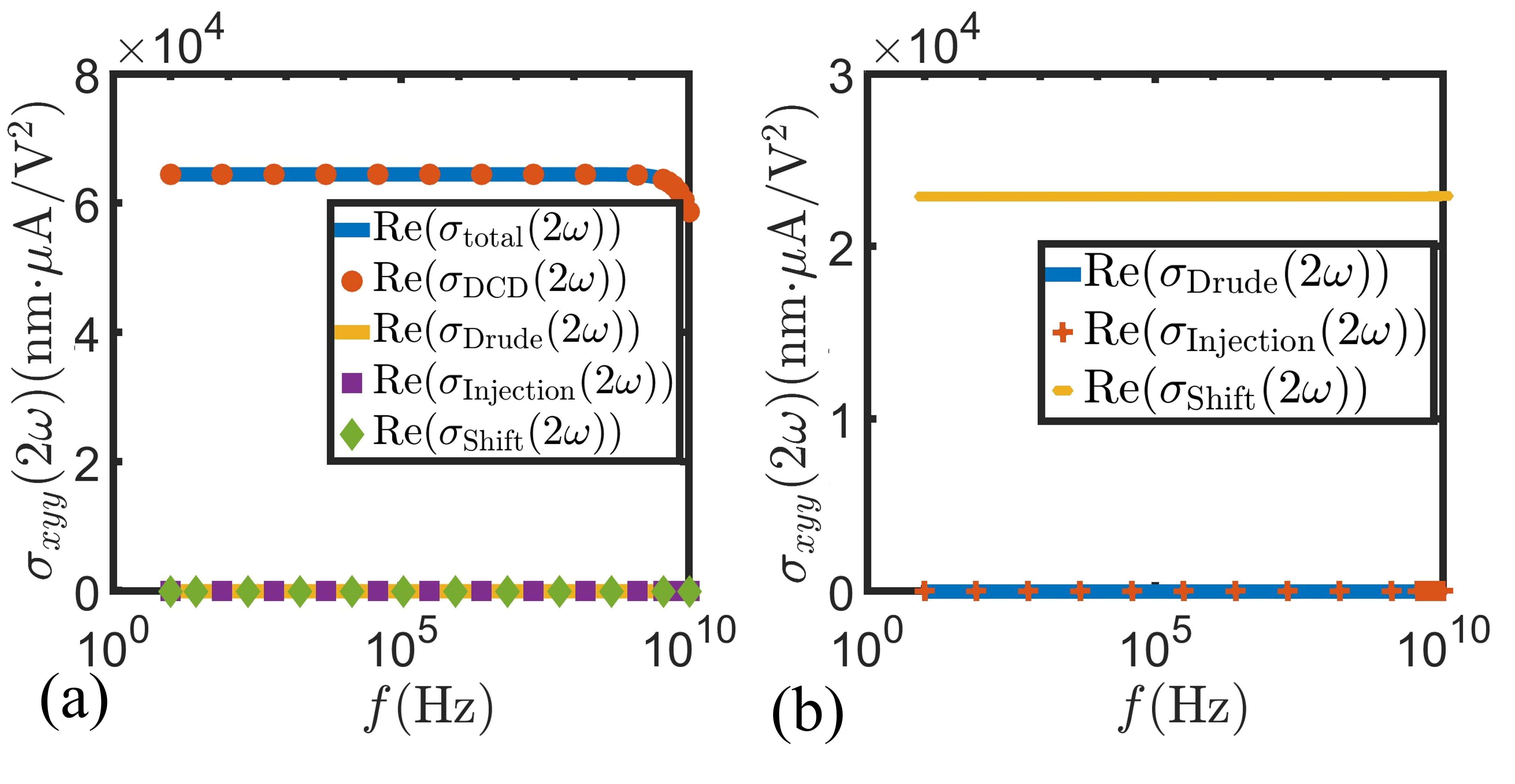}\\
\caption{Nonlinear conductivities with $T$ = 80 K, $\tau$ = 5 ps, $E_f = 0.12$ eV. (a) The real distribution of nonlinear electrical conductivity $\sigma_{xyy}$ varies with frequency (b) An enlarged view of the conductivity of Drude, Injection, and Shift in (a).	}
\label{fig_sigma}
\end{figure}

The total Kerr angle shown in Fig.~\ref{fig_Kerr_t} is clearly dominated by the BCD contribution, with only minimal impact from the Drude, Shift, and Injection effects. 
This aligns with the findings in Fig.~\ref{fig_sigma}(a), which identifies the BCD effect as the primary driver of the nonlinear response in ML WTe$2$. 
The contributions from the other nonlinear mechanisms are insignificant because they are overshadowed by the large background Kerr effect arising from the crystal's intrinsic anisotropy ($\sigma_{xx} \neq \sigma_{yy}$). 
As a result, including these weaker effects does not noticeably alter the signal from the purely linear case (see Fig.~5 in Ref. \cite{SM}). 
Notably, the dominant BCD Kerr angle exhibits characteristics of SHG over time, with a period of 1/$(2f) = 0.5 $ ms.

Having analyzed the composition of the Kerr angle at a fixed frequency, we now examine its behavior across a range of frequencies.
As shown in Fig.~\ref{fig_sigma}, the nonlinear conductivity shows little variation with frequency. 
Therefore, the NEOKE is also expected to be largely independent of frequency. 
This is corroborated by Figs.~\ref{fig_Kerr_t}(f-j), wherein the temporal evolution of the total Kerr angle remains approximately constant across frequencies. 
Notably, while the variation period decreases with rising frequency, it consistently aligns with the second-harmonic timescale.
%%%%%%

In addition to frequency, the amplitude of the applied electric field is another critical parameter. 
Fig.~\ref{fig_Kerr_E} illustrates the dependence of the total Kerr angle on the electric field intensity at an optical frequency of f = 1 kHz. 
As predicted by Eq.~\ref{Knm}, the NEOKE is significantly impacted by the field strength; a stronger field amplifies the nonlinear contribution, leading to a larger variation in the Kerr angle. 
Specifically, as the electric field strength increases, the amplitude of the Kerr angle grows significantly, while the small offset $\theta_k-\theta_0$ is progressively masked. 
This occurs because, at low field strengths, the nonlinear effect only provides a weak enhancement to the crystal's anisotropy. 
At higher field strengths, the dominant amplification of the Kerr angle overshadows this minor deviation.

%%%%

To understand the origin of these NEOKE and explain why the BCD effect is dominant, we now calculate the magnitudes of the various nonlinear conductivities.
Fig.~\ref{fig_sigma} presents these conductivity components. 
The BCD effect clearly dominates the nonlinear conductivity in the low-frequency regime ($f < 10^9$ Hz). 
This fundamental dominance in conductivity directly explains why the BCD contribution to the Kerr angle is also the largest (detailed conductivity data can be found in Fig. 6 of Ref. \cite{SM}). Therefore, the NEOKE provides a powerful method for measuring BCD-induced nonlinear transport and, in turn, for indirectly probing the Berry curvature dipole itself.

%%%%%%%

This finding suggests a viable experimental approach for detecting nonlinear effects using the time-resolved Kerr effect (TRKE) \cite{PhysRevResearch.6.013107,T-MOKE2}. 
For instance, it should be experimentally feasible to observe that the NEOKE amplitude increases with the amplitude of the applied optical electric field; as confirmed by our analysis, such modifications in the Kerr angle waveform at different field strengths would provide clear evidence of the underlying BCD contribution. 
Furthermore, the overall temporal evolution of the electro-optic Kerr angle can be used to identify the emergence of the BCD effect. 
Our theoretical predictions indicate that the characteristic timescales of this time-dependent NEOKE are on the order of milliseconds. 
Given that experimental techniques can achieve temporal resolutions in the picosecond domain \cite{PhysRevResearch.6.013107,T-MOKE2}, these millisecond-scale dynamics should be readily observable, providing a solid theoretical basis for the experimental pursuit of the time-dependent NEOKE.

%%%%%%%%%%%%%%%%%%%%%%%%%%%%%%%%%%%%%%%%%%%%%%%%%%%%%%%%%%%%%%%%%%%%%%%%%%%%%%%%%%%%%%%%%%%%%%%%%%%%%%%%%%%%%%%%
%%%%%%%%%%%%%%%%%%%%%%%%%%%%%%%%%%%%%%%%%%%%%%%%%%%%%%%%%%%%%%%%%%%%%%%%%%%%%%%%%%%%%%%%%%%%%%%%%%%%%%%%%%%%%%%%

\section{Conclusion}

In this work, we extend the concept of the MOKE to the NEOKE in non-magnetic materials, presenting a theoretical investigation of the NEOKE in monolayer WTe$_2$. 
We find that the material's nonlinear conductivity gives rise to a time-dependent Kerr angle. 
Our calculations reveal that this NEOKE is overwhelmingly dominated by the BCD across a wide range of optical frequencies and electric field strengths, and we detail the distinct temporal and field-dependent characteristics of this BCD-driven effect.

These findings suggest a viable experimental approach for detecting the predicted nonlinear effects using the TRKE. 
Our theory predicts two distinct, observable signatures. 
First, the emergence of the BCD effect can be identified by the temporal evolution of the Kerr angle, which exhibits characteristic dynamics on the order of milliseconds—a timescale readily observable with current TRKE techniques. 
Second, modifications in the Kerr angle waveform at different field strengths, specifically an increase in amplitude with the applied optical field, would provide further evidence of the underlying nonlinear contributions. 

The combination of these unique temporal and field-dependent signatures provides a robust basis for the experimental pursuit of the NEOKE in monolayer WTe$_2$. 
More broadly, our work proposes the time-resolved NEOKE as a novel, all-optical probe for detecting nonlinear transport and exploring the topological properties of quantum materials.

\section {Acknowledgements}
This work is supported by %
G.S. is supported in part by the Innovation Program for Quantum Science and Technology
(Grant No. 2024ZD0300500), NSFC (Grant No. 12447101) and CAS.


\begin{thebibliography}{52}%
\makeatletter
\providecommand \@ifxundefined [1]{%
 \@ifx{#1\undefined}
}%
\providecommand \@ifnum [1]{%
 \ifnum #1\expandafter \@firstoftwo
 \else \expandafter \@secondoftwo
 \fi
}%
\providecommand \@ifx [1]{%
 \ifx #1\expandafter \@firstoftwo
 \else \expandafter \@secondoftwo
 \fi
}%
\providecommand \natexlab [1]{#1}%
\providecommand \enquote  [1]{``#1''}%
\providecommand \bibnamefont  [1]{#1}%
\providecommand \bibfnamefont [1]{#1}%
\providecommand \citenamefont [1]{#1}%
\providecommand \href@noop [0]{\@secondoftwo}%
\providecommand \href [0]{\begingroup \@sanitize@url \@href}%
\providecommand \@href[1]{\@@startlink{#1}\@@href}%
\providecommand \@@href[1]{\endgroup#1\@@endlink}%
\providecommand \@sanitize@url [0]{\catcode `\\12\catcode `\$12\catcode `\&12\catcode `\#12\catcode `\^12\catcode `\_12\catcode `\%12\relax}%
\providecommand \@@startlink[1]{}%
\providecommand \@@endlink[0]{}%
\providecommand \url  [0]{\begingroup\@sanitize@url \@url }%
\providecommand \@url [1]{\endgroup\@href {#1}{\urlprefix }}%
\providecommand \urlprefix  [0]{URL }%
\providecommand \Eprint [0]{\href }%
\providecommand \doibase [0]{https://doi.org/}%
\providecommand \selectlanguage [0]{\@gobble}%
\providecommand \bibinfo  [0]{\@secondoftwo}%
\providecommand \bibfield  [0]{\@secondoftwo}%
\providecommand \translation [1]{[#1]}%
\providecommand \BibitemOpen [0]{}%
\providecommand \bibitemStop [0]{}%
\providecommand \bibitemNoStop [0]{.\EOS\space}%
\providecommand \EOS [0]{\spacefactor3000\relax}%
\providecommand \BibitemShut  [1]{\csname bibitem#1\endcsname}%
\let\auto@bib@innerbib\@empty
%</preamble>
\bibitem [{\citenamefont {Gong}\ \emph {et~al.}(2025)\citenamefont {Gong}, \citenamefont {Du}, \citenamefont {Sun}, \citenamefont {Lu},\ and\ \citenamefont {Xie}}]{gong2025nonlineartransporttheoryorder}%
  \BibitemOpen
  \bibfield  {author} {\bibinfo {author} {\bibfnamefont {Z.-H.}\ \bibnamefont {Gong}}, \bibinfo {author} {\bibfnamefont {Z.~Z.}\ \bibnamefont {Du}}, \bibinfo {author} {\bibfnamefont {H.-P.}\ \bibnamefont {Sun}}, \bibinfo {author} {\bibfnamefont {H.-Z.}\ \bibnamefont {Lu}},\ and\ \bibinfo {author} {\bibfnamefont {X.~C.}\ \bibnamefont {Xie}},\ } {\bibinfo {title} {Nonlinear transport theory at the order of quantum metric}} (\bibinfo {year} {2025}),\ \Eprint {https://arxiv.org/abs/2410.04995} {arXiv:2410.04995} \BibitemShut {NoStop}%
\bibitem [{\citenamefont {Sodemann}\ and\ \citenamefont {Fu}(2015{\natexlab{a}})}]{PhysRevLett.115.216806}%
  \BibitemOpen
  \bibfield  {author} {\bibinfo {author} {\bibfnamefont {I.}~\bibnamefont {Sodemann}}\ and\ \bibinfo {author} {\bibfnamefont {L.}~\bibnamefont {Fu}},\ }\bibfield  {title} {\bibinfo {title} {Quantum nonlinear hall effect induced by berry curvature dipole in time-reversal invariant materials},\ }\href {https://doi.org/10.1103/PhysRevLett.115.216806} {\bibfield  {journal} {\bibinfo  {journal} {Phys. Rev. Lett.}\ }\textbf {\bibinfo {volume} {115}},\ \bibinfo {pages} {216806} (\bibinfo {year} {2015}{\natexlab{a}})}\BibitemShut {NoStop}%
\bibitem [{\citenamefont {Low}\ \emph {et~al.}(2015)\citenamefont {Low}, \citenamefont {Jiang},\ and\ \citenamefont {Guinea}}]{PhysRevB.92.235447}%
  \BibitemOpen
  \bibfield  {author} {\bibinfo {author} {\bibfnamefont {T.}~\bibnamefont {Low}}, \bibinfo {author} {\bibfnamefont {Y.}~\bibnamefont {Jiang}},\ and\ \bibinfo {author} {\bibfnamefont {F.}~\bibnamefont {Guinea}},\ }\bibfield  {title} {\bibinfo {title} {Topological currents in black phosphorus with broken inversion symmetry},\ }\href {https://doi.org/10.1103/PhysRevB.92.235447} {\bibfield  {journal} {\bibinfo  {journal} {Phys. Rev. B}\ }\textbf {\bibinfo {volume} {92}},\ \bibinfo {pages} {235447} (\bibinfo {year} {2015})}\BibitemShut {NoStop}%
\bibitem [{\citenamefont {Ma}\ \emph {et~al.}(2019)\citenamefont {Ma}, \citenamefont {Xu}, \citenamefont {Shen}, \citenamefont {MacNeill}, \citenamefont {Fatemi}, \citenamefont {Chang}, \citenamefont {Mier~Valdivia}, \citenamefont {Wu}, \citenamefont {Du}, \citenamefont {Hsu} \emph {et~al.}}]{ma2019observation}%
  \BibitemOpen
  \bibfield  {author} {\bibinfo {author} {\bibfnamefont {Q.}~\bibnamefont {Ma}}, \bibinfo {author} {\bibfnamefont {S.-Y.}\ \bibnamefont {Xu}}, \bibinfo {author} {\bibfnamefont {H.}~\bibnamefont {Shen}}, \bibinfo {author} {\bibfnamefont {D.}~\bibnamefont {MacNeill}}, \bibinfo {author} {\bibfnamefont {V.}~\bibnamefont {Fatemi}}, \bibinfo {author} {\bibfnamefont {T.-R.}\ \bibnamefont {Chang}}, \bibinfo {author} {\bibfnamefont {A.~M.}\ \bibnamefont {Mier~Valdivia}}, \bibinfo {author} {\bibfnamefont {S.}~\bibnamefont {Wu}}, \bibinfo {author} {\bibfnamefont {Z.}~\bibnamefont {Du}}, \bibinfo {author} {\bibfnamefont {C.-H.}\ \bibnamefont {Hsu}}, \emph {et~al.},\ }\bibfield  {title} {\bibinfo {title} {Observation of the nonlinear hall effect under time-reversal-symmetric conditions},\ }\href {https://doi.org/10.1038/s41586-018-0807-6} {\bibfield  {journal} {\bibinfo  {journal} {Nature}\ }\textbf {\bibinfo {volume} {565}},\ \bibinfo {pages} {337} (\bibinfo {year} {2019})}\BibitemShut {NoStop}%
\bibitem [{\citenamefont {Wang}\ \emph {et~al.}(2024)\citenamefont {Wang}, \citenamefont {Zhang}, \citenamefont {Zhu},\ and\ \citenamefont {Su}}]{PhysRevB.109.085419}%
  \BibitemOpen
  \bibfield  {author} {\bibinfo {author} {\bibfnamefont {Y.}~\bibnamefont {Wang}}, \bibinfo {author} {\bibfnamefont {Z.}~\bibnamefont {Zhang}}, \bibinfo {author} {\bibfnamefont {Z.-G.}\ \bibnamefont {Zhu}},\ and\ \bibinfo {author} {\bibfnamefont {G.}~\bibnamefont {Su}},\ }\bibfield  {title} {\bibinfo {title} {Intrinsic nonlinear ohmic current},\ }\href {https://doi.org/10.1103/PhysRevB.109.085419} {\bibfield  {journal} {\bibinfo  {journal} {Phys. Rev. B}\ }\textbf {\bibinfo {volume} {109}},\ \bibinfo {pages} {085419} (\bibinfo {year} {2024})}\BibitemShut {NoStop}%
\bibitem [{\citenamefont {Du}\ \emph {et~al.}(2019)\citenamefont {Du}, \citenamefont {Wang}, \citenamefont {Li}, \citenamefont {Lu},\ and\ \citenamefont {Xie}}]{du2019disorder}%
  \BibitemOpen
  \bibfield  {author} {\bibinfo {author} {\bibfnamefont {Z.}~\bibnamefont {Du}}, \bibinfo {author} {\bibfnamefont {C.}~\bibnamefont {Wang}}, \bibinfo {author} {\bibfnamefont {S.}~\bibnamefont {Li}}, \bibinfo {author} {\bibfnamefont {H.-Z.}\ \bibnamefont {Lu}},\ and\ \bibinfo {author} {\bibfnamefont {X.}~\bibnamefont {Xie}},\ }\bibfield  {title} {\bibinfo {title} {Disorder-induced nonlinear hall effect with time-reversal symmetry},\ }\href@noop {} {\bibfield  {journal} {\bibinfo  {journal} {Nature communications}\ }\textbf {\bibinfo {volume} {10}},\ \bibinfo {pages} {3047} (\bibinfo {year} {2019})}\BibitemShut {NoStop}%
\bibitem [{\citenamefont {Isobe}\ \emph {et~al.}(2020)\citenamefont {Isobe}, \citenamefont {Xu},\ and\ \citenamefont {Fu}}]{doi:10.1126/sciadv.aay2497}%
  \BibitemOpen
  \bibfield  {author} {\bibinfo {author} {\bibfnamefont {H.}~\bibnamefont {Isobe}}, \bibinfo {author} {\bibfnamefont {S.-Y.}\ \bibnamefont {Xu}},\ and\ \bibinfo {author} {\bibfnamefont {L.}~\bibnamefont {Fu}},\ }\bibfield  {title} {\bibinfo {title} {High-frequency rectification via chiral bloch electrons},\ }\href {https://doi.org/10.1126/sciadv.aay2497} {\bibfield  {journal} {\bibinfo  {journal} {Science Advances}\ }\textbf {\bibinfo {volume} {6}},\ \bibinfo {pages} {eaay2497} (\bibinfo {year} {2020})},\  \BibitemShut {NoStop}%
\bibitem [{\citenamefont {Du}\ \emph {et~al.}(2021{\natexlab{a}})\citenamefont {Du}, \citenamefont {Wang}, \citenamefont {Sun}, \citenamefont {Lu},\ and\ \citenamefont {Xie}}]{du2021quantum}%
  \BibitemOpen
  \bibfield  {author} {\bibinfo {author} {\bibfnamefont {Z.}~\bibnamefont {Du}}, \bibinfo {author} {\bibfnamefont {C.}~\bibnamefont {Wang}}, \bibinfo {author} {\bibfnamefont {H.-P.}\ \bibnamefont {Sun}}, \bibinfo {author} {\bibfnamefont {H.-Z.}\ \bibnamefont {Lu}},\ and\ \bibinfo {author} {\bibfnamefont {X.}~\bibnamefont {Xie}},\ }\bibfield  {title} {\bibinfo {title} {Quantum theory of the nonlinear hall effect},\ }\href {https://www.nature.com/articles/s41467-019-10941-3} {\bibfield  {journal} {\bibinfo  {journal} {Nature communications}\ }\textbf {\bibinfo {volume} {12}},\ \bibinfo {pages} {5038} (\bibinfo {year} {2021}{\natexlab{a}})}\BibitemShut {NoStop}%
\bibitem [{\citenamefont {Ho}\ \emph {et~al.}(2021)\citenamefont {Ho}, \citenamefont {Chang}, \citenamefont {Hsieh}, \citenamefont {Lo}, \citenamefont {Huang}, \citenamefont {Vu}, \citenamefont {Ortix},\ and\ \citenamefont {Chen}}]{ho2021hall}%
  \BibitemOpen
  \bibfield  {author} {\bibinfo {author} {\bibfnamefont {S.-C.}\ \bibnamefont {Ho}}, \bibinfo {author} {\bibfnamefont {C.-H.}\ \bibnamefont {Chang}}, \bibinfo {author} {\bibfnamefont {Y.-C.}\ \bibnamefont {Hsieh}}, \bibinfo {author} {\bibfnamefont {S.-T.}\ \bibnamefont {Lo}}, \bibinfo {author} {\bibfnamefont {B.}~\bibnamefont {Huang}}, \bibinfo {author} {\bibfnamefont {T.-H.-Y.}\ \bibnamefont {Vu}}, \bibinfo {author} {\bibfnamefont {C.}~\bibnamefont {Ortix}},\ and\ \bibinfo {author} {\bibfnamefont {T.-M.}\ \bibnamefont {Chen}},\ }\bibfield  {title} {\bibinfo {title} {Hall effects in artificially corrugated bilayer graphene without breaking time-reversal symmetry},\ }\href {https://www.nature.com/articles/s41928-021-00537-5} {\bibfield  {journal} {\bibinfo  {journal} {Nature Electronics}\ }\textbf {\bibinfo {volume} {4}},\ \bibinfo {pages} {116} (\bibinfo {year} {2021})}\BibitemShut {NoStop}%
\bibitem [{\citenamefont {Watanabe}\ and\ \citenamefont {Yanase}(2020)}]{PhysRevResearch.2.043081}%
  \BibitemOpen
  \bibfield  {author} {\bibinfo {author} {\bibfnamefont {H.}~\bibnamefont {Watanabe}}\ and\ \bibinfo {author} {\bibfnamefont {Y.}~\bibnamefont {Yanase}},\ }\bibfield  {title} {\bibinfo {title} {Nonlinear electric transport in odd-parity magnetic multipole systems: Application to mn-based compounds},\ }\href {https://doi.org/10.1103/PhysRevResearch.2.043081} {\bibfield  {journal} {\bibinfo  {journal} {Phys. Rev. Res.}\ }\textbf {\bibinfo {volume} {2}},\ \bibinfo {pages} {043081} (\bibinfo {year} {2020})}\BibitemShut {NoStop}%
\bibitem [{\citenamefont {Duan}\ \emph {et~al.}(2022)\citenamefont {Duan}, \citenamefont {Jian}, \citenamefont {Gao}, \citenamefont {Peng}, \citenamefont {Zhong}, \citenamefont {Feng}, \citenamefont {Mao},\ and\ \citenamefont {Yao}}]{PhysRevLett.129.186801}%
  \BibitemOpen
  \bibfield  {author} {\bibinfo {author} {\bibfnamefont {J.}~\bibnamefont {Duan}}, \bibinfo {author} {\bibfnamefont {Y.}~\bibnamefont {Jian}}, \bibinfo {author} {\bibfnamefont {Y.}~\bibnamefont {Gao}}, \bibinfo {author} {\bibfnamefont {H.}~\bibnamefont {Peng}}, \bibinfo {author} {\bibfnamefont {J.}~\bibnamefont {Zhong}}, \bibinfo {author} {\bibfnamefont {Q.}~\bibnamefont {Feng}}, \bibinfo {author} {\bibfnamefont {J.}~\bibnamefont {Mao}},\ and\ \bibinfo {author} {\bibfnamefont {Y.}~\bibnamefont {Yao}},\ }\bibfield  {title} {\bibinfo {title} {Giant second-order nonlinear hall effect in twisted bilayer graphene},\ }\href {https://doi.org/10.1103/PhysRevLett.129.186801} {\bibfield  {journal} {\bibinfo  {journal} {Phys. Rev. Lett.}\ }\textbf {\bibinfo {volume} {129}},\ \bibinfo {pages} {186801} (\bibinfo {year} {2022})}\BibitemShut {NoStop}%
\bibitem [{\citenamefont {He}\ \emph {et~al.}(2024)\citenamefont {He}, \citenamefont {Isobe}, \citenamefont {Koon}, \citenamefont {Tan}, \citenamefont {Hu}, \citenamefont {Li}, \citenamefont {Nagaosa},\ and\ \citenamefont {Shen}}]{he2024third}%
  \BibitemOpen
  \bibfield  {author} {\bibinfo {author} {\bibfnamefont {P.}~\bibnamefont {He}}, \bibinfo {author} {\bibfnamefont {H.}~\bibnamefont {Isobe}}, \bibinfo {author} {\bibfnamefont {G.~K.~W.}\ \bibnamefont {Koon}}, \bibinfo {author} {\bibfnamefont {J.~Y.}\ \bibnamefont {Tan}}, \bibinfo {author} {\bibfnamefont {J.}~\bibnamefont {Hu}}, \bibinfo {author} {\bibfnamefont {J.}~\bibnamefont {Li}}, \bibinfo {author} {\bibfnamefont {N.}~\bibnamefont {Nagaosa}},\ and\ \bibinfo {author} {\bibfnamefont {J.}~\bibnamefont {Shen}},\ }\bibfield  {title} {\bibinfo {title} {Third-order nonlinear hall effect in a quantum hall system},\ }\href {https://www.nature.com/articles/s41565-024-01730-1} {\bibfield  {journal} {\bibinfo  {journal} {Nature Nanotechnology}\ }\textbf {\bibinfo {volume} {19}},\ \bibinfo {pages} {1460} (\bibinfo {year} {2024})}\BibitemShut {NoStop}%
\bibitem [{\citenamefont {Yao}\ \emph {et~al.}(2024)\citenamefont {Yao}, \citenamefont {Liu},\ and\ \citenamefont {Duan}}]{PhysRevB.110.115123}%
  \BibitemOpen
  \bibfield  {author} {\bibinfo {author} {\bibfnamefont {J.}~\bibnamefont {Yao}}, \bibinfo {author} {\bibfnamefont {Y.}~\bibnamefont {Liu}},\ and\ \bibinfo {author} {\bibfnamefont {W.}~\bibnamefont {Duan}},\ }\bibfield  {title} {\bibinfo {title} {Geometrical nonlinear hall effect induced by lorentz force},\ }\href {https://doi.org/10.1103/PhysRevB.110.115123} {\bibfield  {journal} {\bibinfo  {journal} {Phys. Rev. B}\ }\textbf {\bibinfo {volume} {110}},\ \bibinfo {pages} {115123} (\bibinfo {year} {2024})}\BibitemShut {NoStop}%
\bibitem [{\citenamefont {Chen}\ \emph {et~al.}(2024)\citenamefont {Chen}, \citenamefont {Zhai}, \citenamefont {Xiao},\ and\ \citenamefont {Yao}}]{PhysRevResearch.6.L012059}%
  \BibitemOpen
  \bibfield  {author} {\bibinfo {author} {\bibfnamefont {C.}~\bibnamefont {Chen}}, \bibinfo {author} {\bibfnamefont {D.}~\bibnamefont {Zhai}}, \bibinfo {author} {\bibfnamefont {C.}~\bibnamefont {Xiao}},\ and\ \bibinfo {author} {\bibfnamefont {W.}~\bibnamefont {Yao}},\ }\bibfield  {title} {\bibinfo {title} {Crossed nonlinear dynamical hall effect in twisted bilayers},\ }\href {https://doi.org/10.1103/PhysRevResearch.6.L012059} {\bibfield  {journal} {\bibinfo  {journal} {Phys. Rev. Res.}\ }\textbf {\bibinfo {volume} {6}},\ \bibinfo {pages} {L012059} (\bibinfo {year} {2024})}\BibitemShut {NoStop}%
\bibitem [{\citenamefont {Xiang}\ and\ \citenamefont {Wang}(2024)}]{PhysRevB.109.075419}%
  \BibitemOpen
  \bibfield  {author} {\bibinfo {author} {\bibfnamefont {L.}~\bibnamefont {Xiang}}\ and\ \bibinfo {author} {\bibfnamefont {J.}~\bibnamefont {Wang}},\ }\bibfield  {title} {\bibinfo {title} {Intrinsic in-plane magnetononlinear hall effect in tilted weyl semimetals},\ }\href {https://doi.org/10.1103/PhysRevB.109.075419} {\bibfield  {journal} {\bibinfo  {journal} {Phys. Rev. B}\ }\textbf {\bibinfo {volume} {109}},\ \bibinfo {pages} {075419} (\bibinfo {year} {2024})}\BibitemShut {NoStop}%
\bibitem [{\citenamefont {Xiong}\ \emph {et~al.}(2025)\citenamefont {Xiong}, \citenamefont {Gong},\ and\ \citenamefont {Jin}}]{xiong2025strain}%
  \BibitemOpen
  \bibfield  {author} {\bibinfo {author} {\bibfnamefont {Y.}~\bibnamefont {Xiong}}, \bibinfo {author} {\bibfnamefont {Z.}~\bibnamefont {Gong}},\ and\ \bibinfo {author} {\bibfnamefont {H.}~\bibnamefont {Jin}},\ }\bibfield  {title} {\bibinfo {title} {Strain tuning of the nonlinear anomalous hall effect in mos2 monolayer},\ }\href {https://iopscience.iop.org/article/10.1088/1361-648X/adc64a} {\bibfield  {journal} {\bibinfo  {journal} {Journal of Physics: Condensed Matter}\ }\textbf {\bibinfo {volume} {37}},\ \bibinfo {pages} {195301} (\bibinfo {year} {2025})}\BibitemShut {NoStop}%
\bibitem [{\citenamefont {Zhong}\ \emph {et~al.}(2019)\citenamefont {Zhong}, \citenamefont {Feng}, \citenamefont {Liu}, \citenamefont {Zhang},\ and\ \citenamefont {Cao}}]{ZHONG201981}%
  \BibitemOpen
  \bibfield  {author} {\bibinfo {author} {\bibfnamefont {Y.}~\bibnamefont {Zhong}}, \bibinfo {author} {\bibfnamefont {W.}~\bibnamefont {Feng}}, \bibinfo {author} {\bibfnamefont {Z.}~\bibnamefont {Liu}}, \bibinfo {author} {\bibfnamefont {C.}~\bibnamefont {Zhang}},\ and\ \bibinfo {author} {\bibfnamefont {J.}~\bibnamefont {Cao}},\ }\bibfield  {title} {\bibinfo {title} {Nonlinear optical conductivity of weyl semimetals in the terahertz regime},\ }\href {https://doi.org/https://doi.org/10.1016/j.physb.2018.11.051} {\bibfield  {journal} {\bibinfo  {journal} {Physica B: Condensed Matter}\ }\textbf {\bibinfo {volume} {555}},\ \bibinfo {pages} {81} (\bibinfo {year} {2019})}\BibitemShut {NoStop}%
\bibitem [{\citenamefont {Choi}\ \emph {et~al.}(2021{\natexlab{a}})\citenamefont {Choi}, \citenamefont {Doan}, \citenamefont {Kim},\ and\ \citenamefont {Choi}}]{choi2021nonlinearopticalhalleffect}%
  \BibitemOpen
  \bibfield  {author} {\bibinfo {author} {\bibfnamefont {Y.-G.}\ \bibnamefont {Choi}}, \bibinfo {author} {\bibfnamefont {M.-H.}\ \bibnamefont {Doan}}, \bibinfo {author} {\bibfnamefont {Y.}~\bibnamefont {Kim}},\ and\ \bibinfo {author} {\bibfnamefont {G.-M.}\ \bibnamefont {Choi}},\ } {\bibinfo {title} {Nonlinear optical hall effect in weyl semimetal wte2}} (\bibinfo {year} {2021}{\natexlab{a}}),\ \Eprint {https://arxiv.org/abs/2103.08173} {arXiv:2103.08173} \BibitemShut {NoStop}%
\bibitem [{\citenamefont {Zhang}\ \emph {et~al.}(2019)\citenamefont {Zhang}, \citenamefont {Ooi}, \citenamefont {Chen}, \citenamefont {Ang},\ and\ \citenamefont {Sin~Ang}}]{zhang2019optical}%
  \BibitemOpen
  \bibfield  {author} {\bibinfo {author} {\bibfnamefont {T.}~\bibnamefont {Zhang}}, \bibinfo {author} {\bibfnamefont {K.}~\bibnamefont {Ooi}}, \bibinfo {author} {\bibfnamefont {W.}~\bibnamefont {Chen}}, \bibinfo {author} {\bibfnamefont {L.}~\bibnamefont {Ang}},\ and\ \bibinfo {author} {\bibfnamefont {Y.}~\bibnamefont {Sin~Ang}},\ }\bibfield  {title} {\bibinfo {title} {Optical kerr effect and third harmonic generation in topological dirac/weyl semimetal},\ }\href {https://pubmed.ncbi.nlm.nih.gov/31878597/} {\bibfield  {journal} {\bibinfo  {journal} {Optics express}\ }\textbf {\bibinfo {volume} {27}},\ \bibinfo {pages} {38270} (\bibinfo {year} {2019})}\BibitemShut {NoStop}%
\bibitem [{\citenamefont {Morimoto}\ and\ \citenamefont {Nagaosa}(2016)}]{doi:10.1126/sciadv.1501524}%
  \BibitemOpen
  \bibfield  {author} {\bibinfo {author} {\bibfnamefont {T.}~\bibnamefont {Morimoto}}\ and\ \bibinfo {author} {\bibfnamefont {N.}~\bibnamefont {Nagaosa}},\ }\bibfield  {title} {\bibinfo {title} {Topological nature of nonlinear optical effects in solids},\ }\href {https://doi.org/10.1126/sciadv.1501524} {\bibfield  {journal} {\bibinfo  {journal} {Science Advances}\ }\textbf {\bibinfo {volume} {2}},\ \bibinfo {pages} {e1501524} (\bibinfo {year} {2016})},\  \BibitemShut {NoStop}%
\bibitem [{\citenamefont {Xiao}\ \emph {et~al.}(2010)\citenamefont {Xiao}, \citenamefont {Chang},\ and\ \citenamefont {Niu}}]{RevModPhys.82.1959}%
  \BibitemOpen
  \bibfield  {author} {\bibinfo {author} {\bibfnamefont {D.}~\bibnamefont {Xiao}}, \bibinfo {author} {\bibfnamefont {M.-C.}\ \bibnamefont {Chang}},\ and\ \bibinfo {author} {\bibfnamefont {Q.}~\bibnamefont {Niu}},\ }\bibfield  {title} {\bibinfo {title} {Berry phase effects on electronic properties},\ }\href {https://doi.org/10.1103/RevModPhys.82.1959} {\bibfield  {journal} {\bibinfo  {journal} {Rev. Mod. Phys.}\ }\textbf {\bibinfo {volume} {82}},\ \bibinfo {pages} {1959} (\bibinfo {year} {2010})}\BibitemShut {NoStop}%
\bibitem [{\citenamefont {Nagaosa}\ \emph {et~al.}(2010)\citenamefont {Nagaosa}, \citenamefont {Sinova}, \citenamefont {Onoda}, \citenamefont {MacDonald},\ and\ \citenamefont {Ong}}]{RevModPhys.82.1539}%
  \BibitemOpen
  \bibfield  {author} {\bibinfo {author} {\bibfnamefont {N.}~\bibnamefont {Nagaosa}}, \bibinfo {author} {\bibfnamefont {J.}~\bibnamefont {Sinova}}, \bibinfo {author} {\bibfnamefont {S.}~\bibnamefont {Onoda}}, \bibinfo {author} {\bibfnamefont {A.~H.}\ \bibnamefont {MacDonald}},\ and\ \bibinfo {author} {\bibfnamefont {N.~P.}\ \bibnamefont {Ong}},\ }\bibfield  {title} {\bibinfo {title} {Anomalous hall effect},\ }\href {https://doi.org/10.1103/RevModPhys.82.1539} {\bibfield  {journal} {\bibinfo  {journal} {Rev. Mod. Phys.}\ }\textbf {\bibinfo {volume} {82}},\ \bibinfo {pages} {1539} (\bibinfo {year} {2010})}\BibitemShut {NoStop}%
\bibitem [{\citenamefont {Nakatsuji}\ \emph {et~al.}(2015)\citenamefont {Nakatsuji}, \citenamefont {Kiyohara},\ and\ \citenamefont {Higo}}]{nakatsuji2015large}%
  \BibitemOpen
  \bibfield  {author} {\bibinfo {author} {\bibfnamefont {S.}~\bibnamefont {Nakatsuji}}, \bibinfo {author} {\bibfnamefont {N.}~\bibnamefont {Kiyohara}},\ and\ \bibinfo {author} {\bibfnamefont {T.}~\bibnamefont {Higo}},\ }\bibfield  {title} {\bibinfo {title} {Large anomalous hall effect in a non-collinear antiferromagnet at room temperature},\ }\href {https://www.nature.com/articles/nature15723} {\bibfield  {journal} {\bibinfo  {journal} {Nature}\ }\textbf {\bibinfo {volume} {527}},\ \bibinfo {pages} {212} (\bibinfo {year} {2015})}\BibitemShut {NoStop}%
\bibitem [{\citenamefont {Machida}\ \emph {et~al.}(2010)\citenamefont {Machida}, \citenamefont {Nakatsuji}, \citenamefont {Onoda}, \citenamefont {Tayama},\ and\ \citenamefont {Sakakibara}}]{machida2010time}%
  \BibitemOpen
  \bibfield  {author} {\bibinfo {author} {\bibfnamefont {Y.}~\bibnamefont {Machida}}, \bibinfo {author} {\bibfnamefont {S.}~\bibnamefont {Nakatsuji}}, \bibinfo {author} {\bibfnamefont {S.}~\bibnamefont {Onoda}}, \bibinfo {author} {\bibfnamefont {T.}~\bibnamefont {Tayama}},\ and\ \bibinfo {author} {\bibfnamefont {T.}~\bibnamefont {Sakakibara}},\ }\bibfield  {title} {\bibinfo {title} {Time-reversal symmetry breaking and spontaneous hall effect without magnetic dipole order},\ }\href {https://www.nature.com/articles/nature08680} {\bibfield  {journal} {\bibinfo  {journal} {Nature}\ }\textbf {\bibinfo {volume} {463}},\ \bibinfo {pages} {210} (\bibinfo {year} {2010})}\BibitemShut {NoStop}%
\bibitem [{\citenamefont {Yasuda}\ \emph {et~al.}(2016)\citenamefont {Yasuda}, \citenamefont {Wakatsuki}, \citenamefont {Morimoto}, \citenamefont {Yoshimi}, \citenamefont {Tsukazaki}, \citenamefont {Takahashi}, \citenamefont {Ezawa}, \citenamefont {Kawasaki}, \citenamefont {Nagaosa},\ and\ \citenamefont {Tokura}}]{yasuda2016geometric}%
  \BibitemOpen
  \bibfield  {author} {\bibinfo {author} {\bibfnamefont {K.}~\bibnamefont {Yasuda}}, \bibinfo {author} {\bibfnamefont {R.}~\bibnamefont {Wakatsuki}}, \bibinfo {author} {\bibfnamefont {T.}~\bibnamefont {Morimoto}}, \bibinfo {author} {\bibfnamefont {R.}~\bibnamefont {Yoshimi}}, \bibinfo {author} {\bibfnamefont {A.}~\bibnamefont {Tsukazaki}}, \bibinfo {author} {\bibfnamefont {K.}~\bibnamefont {Takahashi}}, \bibinfo {author} {\bibfnamefont {M.}~\bibnamefont {Ezawa}}, \bibinfo {author} {\bibfnamefont {M.}~\bibnamefont {Kawasaki}}, \bibinfo {author} {\bibfnamefont {N.}~\bibnamefont {Nagaosa}},\ and\ \bibinfo {author} {\bibfnamefont {Y.}~\bibnamefont {Tokura}},\ }\bibfield  {title} {\bibinfo {title} {Geometric hall effects in topological insulator heterostructures},\ }\href {https://www.nature.com/articles/nphys3671} {\bibfield  {journal} {\bibinfo  {journal} {Nature Physics}\ }\textbf {\bibinfo {volume} {12}},\ \bibinfo {pages} {555} (\bibinfo {year} {2016})}\BibitemShut {NoStop}%
\bibitem [{\citenamefont {Zhang}\ \emph {et~al.}(2018)\citenamefont {Zhang}, \citenamefont {Sun},\ and\ \citenamefont {Yan}}]{PhysRevB.97.041101}%
  \BibitemOpen
  \bibfield  {author} {\bibinfo {author} {\bibfnamefont {Y.}~\bibnamefont {Zhang}}, \bibinfo {author} {\bibfnamefont {Y.}~\bibnamefont {Sun}},\ and\ \bibinfo {author} {\bibfnamefont {B.}~\bibnamefont {Yan}},\ }\bibfield  {title} {\bibinfo {title} {Berry curvature dipole in weyl semimetal materials: An ab initio study},\ }\href {https://doi.org/10.1103/PhysRevB.97.041101} {\bibfield  {journal} {\bibinfo  {journal} {Phys. Rev. B}\ }\textbf {\bibinfo {volume} {97}},\ \bibinfo {pages} {041101} (\bibinfo {year} {2018})}\BibitemShut {NoStop}%
\bibitem [{\citenamefont {Singh}\ \emph {et~al.}(2020)\citenamefont {Singh}, \citenamefont {Kim}, \citenamefont {Rabe},\ and\ \citenamefont {Vanderbilt}}]{PhysRevLett.125.046402}%
  \BibitemOpen
  \bibfield  {author} {\bibinfo {author} {\bibfnamefont {S.}~\bibnamefont {Singh}}, \bibinfo {author} {\bibfnamefont {J.}~\bibnamefont {Kim}}, \bibinfo {author} {\bibfnamefont {K.~M.}\ \bibnamefont {Rabe}},\ and\ \bibinfo {author} {\bibfnamefont {D.}~\bibnamefont {Vanderbilt}},\ }\bibfield  {title} {\bibinfo {title} {Engineering weyl phases and nonlinear hall effects in ${\mathrm{t}}_{d}$-${\mathrm{mote}}_{2}$},\ }\href {https://doi.org/10.1103/PhysRevLett.125.046402} {\bibfield  {journal} {\bibinfo  {journal} {Phys. Rev. Lett.}\ }\textbf {\bibinfo {volume} {125}},\ \bibinfo {pages} {046402} (\bibinfo {year} {2020})}\BibitemShut {NoStop}%
\bibitem [{\citenamefont {Facio}\ \emph {et~al.}(2018)\citenamefont {Facio}, \citenamefont {Efremov}, \citenamefont {Koepernik}, \citenamefont {You}, \citenamefont {Sodemann},\ and\ \citenamefont {van~den Brink}}]{PhysRevLett.121.246403}%
  \BibitemOpen
  \bibfield  {author} {\bibinfo {author} {\bibfnamefont {J.~I.}\ \bibnamefont {Facio}}, \bibinfo {author} {\bibfnamefont {D.}~\bibnamefont {Efremov}}, \bibinfo {author} {\bibfnamefont {K.}~\bibnamefont {Koepernik}}, \bibinfo {author} {\bibfnamefont {J.-S.}\ \bibnamefont {You}}, \bibinfo {author} {\bibfnamefont {I.}~\bibnamefont {Sodemann}},\ and\ \bibinfo {author} {\bibfnamefont {J.}~\bibnamefont {van~den Brink}},\ }\bibfield  {title} {\bibinfo {title} {Strongly enhanced berry dipole at topological phase transitions in bitei},\ }\href {https://doi.org/10.1103/PhysRevLett.121.246403} {\bibfield  {journal} {\bibinfo  {journal} {Phys. Rev. Lett.}\ }\textbf {\bibinfo {volume} {121}},\ \bibinfo {pages} {246403} (\bibinfo {year} {2018})}\BibitemShut {NoStop}%
\bibitem [{\citenamefont {Argyres}(1955)}]{Argyres1955}%
  \BibitemOpen
  \bibfield  {author} {\bibinfo {author} {\bibfnamefont {P.~N.}\ \bibnamefont {Argyres}},\ }\bibfield  {title} {\bibinfo {title} {Theory of the faraday and kerr effects in ferromagnetics},\ }\href {https://doi.org/10.1103/PhysRev.97.334} {\bibfield  {journal} {\bibinfo  {journal} {Phys. Rev.}\ }\textbf {\bibinfo {volume} {97}},\ \bibinfo {pages} {334} (\bibinfo {year} {1955})}\BibitemShut {NoStop}%
\bibitem [{\citenamefont {Crova}(1875)}]{Kerr1875}%
  \BibitemOpen
  \bibfield  {author} {\bibinfo {author} {\bibfnamefont {A.}~\bibnamefont {Crova}},\ }\bibfield  {title} {\bibinfo {title} {{J. KERR. -A new relation between electricity and light : dielectrified m{\'e}dia birefringent (Nouvelle relation entre l'{\'e}lectricit{\'e} et la lumi{\`e}re. Bir{\'e}fringence des milieux di{\'e}lectriques transparents); Philosophical Magazine, 3 e s{\'e}rie, t. L, p. 337, novembre 1875}},\ }\href {https://doi.org/10.1051/jphystap:018750040037601} {\bibfield  {journal} {\bibinfo  {journal} {{J. Phys. Theor. Appl.}}\ }\textbf {\bibinfo {volume} {4}},\ \bibinfo {pages} {376} (\bibinfo {year} {1875})}\BibitemShut {NoStop}%
\bibitem [{\citenamefont {Oppeneer}(1999)}]{book}%
  \BibitemOpen
  \bibfield  {author} {\bibinfo {author} {\bibfnamefont {P.}~\bibnamefont {Oppeneer}},\ }\href@noop {} {\emph {\bibinfo {title} {Theory of the Magneto-Optical Kerr Effect in Ferromagnetic Compounds}}}\ (\bibinfo {year} {1999})\BibitemShut {NoStop}%
\bibitem [{\citenamefont {Sodemann}\ and\ \citenamefont {Fu}(2015{\natexlab{b}})}]{sodemann2015quantum}%
  \BibitemOpen
  \bibfield  {author} {\bibinfo {author} {\bibfnamefont {I.}~\bibnamefont {Sodemann}}\ and\ \bibinfo {author} {\bibfnamefont {L.}~\bibnamefont {Fu}},\ }\bibfield  {title} {\bibinfo {title} {Quantum nonlinear hall effect induced by berry curvature dipole in time-reversal invariant materials},\ }\href {https://doi.org/10.1103/PhysRevLett.115.216806} {\bibfield  {journal} {\bibinfo  {journal} {Phys. Rev. Lett.}\ }\textbf {\bibinfo {volume} {115}},\ \bibinfo {pages} {216806} (\bibinfo {year} {2015}{\natexlab{b}})}\BibitemShut {NoStop}%
\bibitem [{\citenamefont {Wang}\ \emph {et~al.}(2023)\citenamefont {Wang}, \citenamefont {Kaplan}, \citenamefont {Zhang}, \citenamefont {Holder}, \citenamefont {Cao}, \citenamefont {Wang}, \citenamefont {Zhou}, \citenamefont {Zhou}, \citenamefont {Jiang}, \citenamefont {Zhang} \emph {et~al.}}]{gao2023}%
  \BibitemOpen
  \bibfield  {author} {\bibinfo {author} {\bibfnamefont {N.}~\bibnamefont {Wang}}, \bibinfo {author} {\bibfnamefont {D.}~\bibnamefont {Kaplan}}, \bibinfo {author} {\bibfnamefont {Z.}~\bibnamefont {Zhang}}, \bibinfo {author} {\bibfnamefont {T.}~\bibnamefont {Holder}}, \bibinfo {author} {\bibfnamefont {N.}~\bibnamefont {Cao}}, \bibinfo {author} {\bibfnamefont {A.}~\bibnamefont {Wang}}, \bibinfo {author} {\bibfnamefont {X.}~\bibnamefont {Zhou}}, \bibinfo {author} {\bibfnamefont {F.}~\bibnamefont {Zhou}}, \bibinfo {author} {\bibfnamefont {Z.}~\bibnamefont {Jiang}}, \bibinfo {author} {\bibfnamefont {C.}~\bibnamefont {Zhang}}, \emph {et~al.},\ }\bibfield  {title} {\bibinfo {title} {Quantum-metric-induced nonlinear transport in a topological antiferromagnet},\ }\href {https://www.nature.com/articles/s41586-023-06363-3} {\bibfield  {journal} {\bibinfo  {journal} {Nature}\ }\textbf {\bibinfo {volume} {621}},\ \bibinfo {pages} {487} (\bibinfo {year} {2023})}\BibitemShut {NoStop}%
\bibitem [{\citenamefont {Du}\ \emph {et~al.}(2021{\natexlab{b}})\citenamefont {Du}, \citenamefont {Lu},\ and\ \citenamefont {Xie}}]{du2021nonlinear}%
  \BibitemOpen
  \bibfield  {author} {\bibinfo {author} {\bibfnamefont {Z.}~\bibnamefont {Du}}, \bibinfo {author} {\bibfnamefont {H.-Z.}\ \bibnamefont {Lu}},\ and\ \bibinfo {author} {\bibfnamefont {X.}~\bibnamefont {Xie}},\ }\bibfield  {title} {\bibinfo {title} {Nonlinear hall effects},\ }\href {https://www.nature.com/articles/s42254-021-00359-6} {\bibfield  {journal} {\bibinfo  {journal} {Nature Reviews Physics}\ }\textbf {\bibinfo {volume} {3}},\ \bibinfo {pages} {744} (\bibinfo {year} {2021}{\natexlab{b}})}\BibitemShut {NoStop}%
\bibitem [{\citenamefont {Kang}\ \emph {et~al.}(2019)\citenamefont {Kang}, \citenamefont {Li}, \citenamefont {Sohn}, \citenamefont {Shan},\ and\ \citenamefont {Mak}}]{kang2019nonlinear}%
  \BibitemOpen
  \bibfield  {author} {\bibinfo {author} {\bibfnamefont {K.}~\bibnamefont {Kang}}, \bibinfo {author} {\bibfnamefont {T.}~\bibnamefont {Li}}, \bibinfo {author} {\bibfnamefont {E.}~\bibnamefont {Sohn}}, \bibinfo {author} {\bibfnamefont {J.}~\bibnamefont {Shan}},\ and\ \bibinfo {author} {\bibfnamefont {K.~F.}\ \bibnamefont {Mak}},\ }\bibfield  {title} {\bibinfo {title} {Nonlinear anomalous hall effect in few-layer wte2},\ }\href {https://doi.org/10.1038/s41563-019-0294-7} {\bibfield  {journal} {\bibinfo  {journal} {Nature materials}\ }\textbf {\bibinfo {volume} {18}},\ \bibinfo {pages} {324} (\bibinfo {year} {2019})}\BibitemShut {NoStop}%
\bibitem [{\citenamefont {Du}\ \emph {et~al.}(2018)\citenamefont {Du}, \citenamefont {Wang}, \citenamefont {Lu},\ and\ \citenamefont {Xie}}]{PhysRevLett.121.266601}%
  \BibitemOpen
  \bibfield  {author} {\bibinfo {author} {\bibfnamefont {Z.~Z.}\ \bibnamefont {Du}}, \bibinfo {author} {\bibfnamefont {C.~M.}\ \bibnamefont {Wang}}, \bibinfo {author} {\bibfnamefont {H.-Z.}\ \bibnamefont {Lu}},\ and\ \bibinfo {author} {\bibfnamefont {X.~C.}\ \bibnamefont {Xie}},\ }\bibfield  {title} {\bibinfo {title} {Band signatures for strong nonlinear hall effect in bilayer ${\mathrm{wte}}_{2}$},\ }\href {https://doi.org/10.1103/PhysRevLett.121.266601} {\bibfield  {journal} {\bibinfo  {journal} {Phys. Rev. Lett.}\ }\textbf {\bibinfo {volume} {121}},\ \bibinfo {pages} {266601} (\bibinfo {year} {2018})}\BibitemShut {NoStop}%
\bibitem [{\citenamefont {Jiang}\ \emph {et~al.}(2015)\citenamefont {Jiang}, \citenamefont {Tang}, \citenamefont {Pan}, \citenamefont {Liu}, \citenamefont {Niu}, \citenamefont {Wang}, \citenamefont {Xu}, \citenamefont {Yang}, \citenamefont {Xie}, \citenamefont {Song}, \citenamefont {Dudin}, \citenamefont {Kim}, \citenamefont {Hoesch}, \citenamefont {Das}, \citenamefont {Vobornik}, \citenamefont {Wan},\ and\ \citenamefont {Feng}}]{PhysRevLett.115.166601}%
  \BibitemOpen
  \bibfield  {author} {\bibinfo {author} {\bibfnamefont {J.}~\bibnamefont {Jiang}}, \bibinfo {author} {\bibfnamefont {F.}~\bibnamefont {Tang}}, \bibinfo {author} {\bibfnamefont {X.~C.}\ \bibnamefont {Pan}}, \bibinfo {author} {\bibfnamefont {H.~M.}\ \bibnamefont {Liu}}, \bibinfo {author} {\bibfnamefont {X.~H.}\ \bibnamefont {Niu}}, \bibinfo {author} {\bibfnamefont {Y.~X.}\ \bibnamefont {Wang}}, \bibinfo {author} {\bibfnamefont {D.~F.}\ \bibnamefont {Xu}}, \bibinfo {author} {\bibfnamefont {H.~F.}\ \bibnamefont {Yang}}, \bibinfo {author} {\bibfnamefont {B.~P.}\ \bibnamefont {Xie}}, \bibinfo {author} {\bibfnamefont {F.~Q.}\ \bibnamefont {Song}}, \bibinfo {author} {\bibfnamefont {P.}~\bibnamefont {Dudin}}, \bibinfo {author} {\bibfnamefont {T.~K.}\ \bibnamefont {Kim}}, \bibinfo {author} {\bibfnamefont {M.}~\bibnamefont {Hoesch}}, \bibinfo {author} {\bibfnamefont {P.~K.}\ \bibnamefont {Das}}, \bibinfo {author} {\bibfnamefont {I.}~\bibnamefont {Vobornik}}, \bibinfo {author} {\bibfnamefont {X.~G.}\ \bibnamefont {Wan}},\ and\ \bibinfo {author} {\bibfnamefont {D.~L.}\ \bibnamefont {Feng}},\ }\bibfield  {title} {\bibinfo {title} {Signature of strong spin-orbital coupling in the large nonsaturating magnetoresistance material ${\mathrm{wte}}_{2}$},\ }\href {https://doi.org/10.1103/PhysRevLett.115.166601} {\bibfield  {journal} {\bibinfo  {journal} {Phys. Rev. Lett.}\ }\textbf {\bibinfo {volume} {115}},\ \bibinfo {pages} {166601} (\bibinfo {year} {2015})}\BibitemShut {NoStop}%
\bibitem [{\citenamefont {Huisman}\ \emph {et~al.}(2017)\citenamefont {Huisman}, \citenamefont {Mikhaylovskiy}, \citenamefont {Rasing}, \citenamefont {Kimel}, \citenamefont {Tsukamoto}, \citenamefont {de~Ronde}, \citenamefont {Ma}, \citenamefont {Fan},\ and\ \citenamefont {Zhou}}]{PhysRevB.95.094418}%
  \BibitemOpen
  \bibfield  {author} {\bibinfo {author} {\bibfnamefont {T.~J.}\ \bibnamefont {Huisman}}, \bibinfo {author} {\bibfnamefont {R.~V.}\ \bibnamefont {Mikhaylovskiy}}, \bibinfo {author} {\bibfnamefont {T.}~\bibnamefont {Rasing}}, \bibinfo {author} {\bibfnamefont {A.~V.}\ \bibnamefont {Kimel}}, \bibinfo {author} {\bibfnamefont {A.}~\bibnamefont {Tsukamoto}}, \bibinfo {author} {\bibfnamefont {B.}~\bibnamefont {de~Ronde}}, \bibinfo {author} {\bibfnamefont {L.}~\bibnamefont {Ma}}, \bibinfo {author} {\bibfnamefont {W.~J.}\ \bibnamefont {Fan}},\ and\ \bibinfo {author} {\bibfnamefont {S.~M.}\ \bibnamefont {Zhou}},\ }\bibfield  {title} {\bibinfo {title} {Sub-100-ps dynamics of the anomalous hall effect at terahertz frequencies},\ }\href {https://doi.org/10.1103/PhysRevB.95.094418} {\bibfield  {journal} {\bibinfo  {journal} {Phys. Rev. B}\ }\textbf {\bibinfo {volume} {95}},\ \bibinfo {pages} {094418} (\bibinfo {year} {2017})}\BibitemShut {NoStop}%
\bibitem [{\citenamefont {Matsyshyn}\ and\ \citenamefont {Sodemann}(2019)}]{matsyshyn2019nonlinear}%
  \BibitemOpen
  \bibfield  {author} {\bibinfo {author} {\bibfnamefont {O.}~\bibnamefont {Matsyshyn}}\ and\ \bibinfo {author} {\bibfnamefont {I.}~\bibnamefont {Sodemann}},\ }\bibfield  {title} {\bibinfo {title} {Nonlinear hall acceleration and the quantum rectification sum rule},\ }\href {https://doi.org/10.1103/PhysRevLett.123.246602} {\bibfield  {journal} {\bibinfo  {journal} {Phys. Rev. Lett.}\ }\textbf {\bibinfo {volume} {123}},\ \bibinfo {pages} {246602} (\bibinfo {year} {2019})}\BibitemShut {NoStop}%
\bibitem [{\citenamefont {Liu}\ \emph {et~al.}(2024)\citenamefont {Liu}, \citenamefont {Zhu},\ and\ \citenamefont {Su}}]{liu2024photogalvanic}%
  \BibitemOpen
  \bibfield  {author} {\bibinfo {author} {\bibfnamefont {Y.}~\bibnamefont {Liu}}, \bibinfo {author} {\bibfnamefont {Z.-G.}\ \bibnamefont {Zhu}},\ and\ \bibinfo {author} {\bibfnamefont {G.}~\bibnamefont {Su}},\ }\bibfield  {title} {\bibinfo {title} {Photogalvanic effect and second-harmonic generation from radio to infrared wavelengths in the wte 2 monolayer},\ }\href {https://doi.org/10.1103/PhysRevB.109.085117} {\bibfield  {journal} {\bibinfo  {journal} {Phys. Rev. B}\ }\textbf {\bibinfo {volume} {109}},\ \bibinfo {pages} {085117} (\bibinfo {year} {2024})}\BibitemShut {NoStop}%
\bibitem [{\citenamefont {Sipe}\ and\ \citenamefont {Shkrebtii}(2000)}]{Sipe2000}%
  \BibitemOpen
  \bibfield  {author} {\bibinfo {author} {\bibfnamefont {J.~E.}\ \bibnamefont {Sipe}}\ and\ \bibinfo {author} {\bibfnamefont {A.~I.}\ \bibnamefont {Shkrebtii}},\ }\bibfield  {title} {\bibinfo {title} {Second-order optical response in semiconductors},\ }\href {https://doi.org/10.1103/PhysRevB.61.5337} {\bibfield  {journal} {\bibinfo  {journal} {Phys. Rev. B}\ }\textbf {\bibinfo {volume} {61}},\ \bibinfo {pages} {5337} (\bibinfo {year} {2000})}\BibitemShut {NoStop}%
\bibitem [{\citenamefont {Antonov}\ \emph {et~al.}(2004)\citenamefont {Antonov}, \citenamefont {Harmon},\ and\ \citenamefont {Yaresko}}]{Antonov2004ElectronicSA}%
  \BibitemOpen
  \bibfield  {author} {\bibinfo {author} {\bibfnamefont {V.~N.}\ \bibnamefont {Antonov}}, \bibinfo {author} {\bibfnamefont {B.~N.}\ \bibnamefont {Harmon}},\ and\ \bibinfo {author} {\bibfnamefont {A.~N.}\ \bibnamefont {Yaresko}},\ }\bibfield  {title} {\bibinfo {title} {Electronic structure and magneto-optical properties of solids}\ }(\bibinfo {year} {2004})\BibitemShut {NoStop}%
\bibitem [{\citenamefont {Kim}\ \emph {et~al.}(2007)\citenamefont {Kim}, \citenamefont {Acbas}, \citenamefont {Yang}, \citenamefont {Ohkubo}, \citenamefont {Christen}, \citenamefont {Mandrus}, \citenamefont {Scarpulla}, \citenamefont {Dubon}, \citenamefont {Schlesinger}, \citenamefont {Khalifah},\ and\ \citenamefont {Cerne}}]{uselinearkerr}%
  \BibitemOpen
  \bibfield  {author} {\bibinfo {author} {\bibfnamefont {M.-H.}\ \bibnamefont {Kim}}, \bibinfo {author} {\bibfnamefont {G.}~\bibnamefont {Acbas}}, \bibinfo {author} {\bibfnamefont {M.-H.}\ \bibnamefont {Yang}}, \bibinfo {author} {\bibfnamefont {I.}~\bibnamefont {Ohkubo}}, \bibinfo {author} {\bibfnamefont {H.}~\bibnamefont {Christen}}, \bibinfo {author} {\bibfnamefont {D.}~\bibnamefont {Mandrus}}, \bibinfo {author} {\bibfnamefont {M.~A.}\ \bibnamefont {Scarpulla}}, \bibinfo {author} {\bibfnamefont {O.~D.}\ \bibnamefont {Dubon}}, \bibinfo {author} {\bibfnamefont {Z.}~\bibnamefont {Schlesinger}}, \bibinfo {author} {\bibfnamefont {P.}~\bibnamefont {Khalifah}},\ and\ \bibinfo {author} {\bibfnamefont {J.}~\bibnamefont {Cerne}},\ }\bibfield  {title} {\bibinfo {title} {Determination of the infrared complex magnetoconductivity tensor in itinerant ferromagnets from faraday and kerr measurements},\ }\href {https://doi.org/10.1103/PhysRevB.75.214416} {\bibfield  {journal} {\bibinfo  {journal} {Phys. Rev. B}\ }\textbf {\bibinfo {volume} {75}},\ \bibinfo {pages} {214416} (\bibinfo {year} {2007})}\BibitemShut {NoStop}%
\bibitem [{\citenamefont {Yang}\ \emph {et~al.}(2022)\citenamefont {Yang}, \citenamefont {Yang}, \citenamefont {Zhou}, \citenamefont {Feng},\ and\ \citenamefont {Yao}}]{yang2022first}%
  \BibitemOpen
  \bibfield  {author} {\bibinfo {author} {\bibfnamefont {X.}~\bibnamefont {Yang}}, \bibinfo {author} {\bibfnamefont {P.}~\bibnamefont {Yang}}, \bibinfo {author} {\bibfnamefont {X.}~\bibnamefont {Zhou}}, \bibinfo {author} {\bibfnamefont {W.}~\bibnamefont {Feng}},\ and\ \bibinfo {author} {\bibfnamefont {Y.}~\bibnamefont {Yao}},\ }\bibfield  {title} {\bibinfo {title} {First-and second-order magneto-optical effects and intrinsically anomalous transport in the two-dimensional van der waals layered magnets cr xy (x= s, se, te; y= cl, br, i)},\ }\href {https://doi.org/10.1103/PhysRevB.106.054408} {\bibfield  {journal} {\bibinfo  {journal} {Phys. Rev. B}\ }\textbf {\bibinfo {volume} {106}},\ \bibinfo {pages} {054408} (\bibinfo {year} {2022})}\BibitemShut {NoStop}%
\bibitem [{\citenamefont {Material}()}]{SM}%
  \BibitemOpen
  \bibfield  {author} {\bibinfo {author} {\bibfnamefont {S.}~\bibnamefont {Material}},\ }\bibfield  {title} {\bibinfo {title} {Supplement material},\ }\href@noop {} {\ }\BibitemShut {NoStop}%
\bibitem [{\citenamefont {Zhao}\ \emph {et~al.}(2021)\citenamefont {Zhao}, \citenamefont {Runburg}, \citenamefont {Fei}, \citenamefont {Mutch}, \citenamefont {Malinowski}, \citenamefont {Sun}, \citenamefont {Huang}, \citenamefont {Pesin}, \citenamefont {Cui}, \citenamefont {Xu}, \citenamefont {Chu},\ and\ \citenamefont {Cobden}}]{WTe2TI}%
  \BibitemOpen
  \bibfield  {author} {\bibinfo {author} {\bibfnamefont {W.}~\bibnamefont {Zhao}}, \bibinfo {author} {\bibfnamefont {E.}~\bibnamefont {Runburg}}, \bibinfo {author} {\bibfnamefont {Z.}~\bibnamefont {Fei}}, \bibinfo {author} {\bibfnamefont {J.}~\bibnamefont {Mutch}}, \bibinfo {author} {\bibfnamefont {P.}~\bibnamefont {Malinowski}}, \bibinfo {author} {\bibfnamefont {B.}~\bibnamefont {Sun}}, \bibinfo {author} {\bibfnamefont {X.}~\bibnamefont {Huang}}, \bibinfo {author} {\bibfnamefont {D.}~\bibnamefont {Pesin}}, \bibinfo {author} {\bibfnamefont {Y.-T.}\ \bibnamefont {Cui}}, \bibinfo {author} {\bibfnamefont {X.}~\bibnamefont {Xu}}, \bibinfo {author} {\bibfnamefont {J.-H.}\ \bibnamefont {Chu}},\ and\ \bibinfo {author} {\bibfnamefont {D.~H.}\ \bibnamefont {Cobden}},\ }\bibfield  {title} {\bibinfo {title} {Determination of the spin axis in quantum spin hall insulator candidate monolayer ${\mathrm{wte}}_{2}$},\ }\href {https://doi.org/10.1103/PhysRevX.11.041034} {\bibfield  {journal} {\bibinfo  {journal} {Phys. Rev. X}\ }\textbf {\bibinfo {volume} {11}},\ \bibinfo {pages} {041034} (\bibinfo {year} {2021})}\BibitemShut {NoStop}%
\bibitem [{\citenamefont {You}\ \emph {et~al.}(2018)\citenamefont {You}, \citenamefont {Fang}, \citenamefont {Xu}, \citenamefont {Kaxiras},\ and\ \citenamefont {Low}}]{WTe2BCD}%
  \BibitemOpen
  \bibfield  {author} {\bibinfo {author} {\bibfnamefont {J.-S.}\ \bibnamefont {You}}, \bibinfo {author} {\bibfnamefont {S.}~\bibnamefont {Fang}}, \bibinfo {author} {\bibfnamefont {S.-Y.}\ \bibnamefont {Xu}}, \bibinfo {author} {\bibfnamefont {E.}~\bibnamefont {Kaxiras}},\ and\ \bibinfo {author} {\bibfnamefont {T.}~\bibnamefont {Low}},\ }\bibfield  {title} {\bibinfo {title} {Berry curvature dipole current in the transition metal dichalcogenides family},\ }\href {https://doi.org/10.1103/PhysRevB.98.121109} {\bibfield  {journal} {\bibinfo  {journal} {Phys. Rev. B}\ }\textbf {\bibinfo {volume} {98}},\ \bibinfo {pages} {121109} (\bibinfo {year} {2018})}\BibitemShut {NoStop}%
\bibitem [{\citenamefont {Shi}\ and\ \citenamefont {Song}(2019)}]{WTE2H}%
  \BibitemOpen
  \bibfield  {author} {\bibinfo {author} {\bibfnamefont {L.-k.}\ \bibnamefont {Shi}}\ and\ \bibinfo {author} {\bibfnamefont {J.~C.~W.}\ \bibnamefont {Song}},\ }\bibfield  {title} {\bibinfo {title} {Symmetry, spin-texture, and tunable quantum geometry in a ${\mathrm{wte}}_{2}$ monolayer},\ }\href {https://doi.org/10.1103/PhysRevB.99.035403} {\bibfield  {journal} {\bibinfo  {journal} {Phys. Rev. B}\ }\textbf {\bibinfo {volume} {99}},\ \bibinfo {pages} {035403} (\bibinfo {year} {2019})}\BibitemShut {NoStop}%
\bibitem [{\citenamefont {Kumar}\ and\ \citenamefont {Ahluwalia}(2012)}]{KUMAR20124627}%
  \BibitemOpen
  \bibfield  {author} {\bibinfo {author} {\bibfnamefont {A.}~\bibnamefont {Kumar}}\ and\ \bibinfo {author} {\bibfnamefont {P.}~\bibnamefont {Ahluwalia}},\ }\bibfield  {title} {\bibinfo {title} {Tunable dielectric response of transition metals dichalcogenides mx2 (m=mo, w; x=s, se, te): Effect of quantum confinement},\ }\href {https://doi.org/https://doi.org/10.1016/j.physb.2012.08.034} {\bibfield  {journal} {\bibinfo  {journal} {Physica B: Condensed Matter}\ }\textbf {\bibinfo {volume} {407}},\ \bibinfo {pages} {4627} (\bibinfo {year} {2012})}\BibitemShut {NoStop}%
\bibitem [{\citenamefont {Probst}\ \emph {et~al.}(2024)\citenamefont {Probst}, \citenamefont {M\"oller}, \citenamefont {Schumacher}, \citenamefont {Brede}, \citenamefont {Dewhurst}, \citenamefont {Reutzel}, \citenamefont {Steil}, \citenamefont {Sharma}, \citenamefont {Jansen},\ and\ \citenamefont {Mathias}}]{PhysRevResearch.6.013107}%
  \BibitemOpen
  \bibfield  {author} {\bibinfo {author} {\bibfnamefont {H.}~\bibnamefont {Probst}}, \bibinfo {author} {\bibfnamefont {C.}~\bibnamefont {M\"oller}}, \bibinfo {author} {\bibfnamefont {M.}~\bibnamefont {Schumacher}}, \bibinfo {author} {\bibfnamefont {T.}~\bibnamefont {Brede}}, \bibinfo {author} {\bibfnamefont {J.~K.}\ \bibnamefont {Dewhurst}}, \bibinfo {author} {\bibfnamefont {M.}~\bibnamefont {Reutzel}}, \bibinfo {author} {\bibfnamefont {D.}~\bibnamefont {Steil}}, \bibinfo {author} {\bibfnamefont {S.}~\bibnamefont {Sharma}}, \bibinfo {author} {\bibfnamefont {G.~S.~M.}\ \bibnamefont {Jansen}},\ and\ \bibinfo {author} {\bibfnamefont {S.}~\bibnamefont {Mathias}},\ }\bibfield  {title} {\bibinfo {title} {Unraveling femtosecond spin and charge dynamics with extreme ultraviolet transverse moke spectroscopy},\ }\href {https://doi.org/10.1103/PhysRevResearch.6.013107} {\bibfield  {journal} {\bibinfo  {journal} {Phys. Rev. Res.}\ }\textbf {\bibinfo {volume} {6}},\ \bibinfo {pages} {013107} (\bibinfo {year} {2024})}\BibitemShut {NoStop}%
\bibitem [{\citenamefont {Anadón}\ \emph {et~al.}(2025)\citenamefont {Anadón}, \citenamefont {Singh}, \citenamefont {Díaz}, \citenamefont {Le-Guen}, \citenamefont {Hohlfeld}, \citenamefont {Wilson}, \citenamefont {Malinowski}, \citenamefont {Hehn},\ and\ \citenamefont {Gorchon}}]{T-MOKE2}%
  \BibitemOpen
  \bibfield  {author} {\bibinfo {author} {\bibfnamefont {A.}~\bibnamefont {Anadón}}, \bibinfo {author} {\bibfnamefont {H.}~\bibnamefont {Singh}}, \bibinfo {author} {\bibfnamefont {E.}~\bibnamefont {Díaz}}, \bibinfo {author} {\bibfnamefont {Y.}~\bibnamefont {Le-Guen}}, \bibinfo {author} {\bibfnamefont {J.}~\bibnamefont {Hohlfeld}}, \bibinfo {author} {\bibfnamefont {R.~B.}\ \bibnamefont {Wilson}}, \bibinfo {author} {\bibfnamefont {G.}~\bibnamefont {Malinowski}}, \bibinfo {author} {\bibfnamefont {M.}~\bibnamefont {Hehn}},\ and\ \bibinfo {author} {\bibfnamefont {J.}~\bibnamefont {Gorchon}},\ } {\bibinfo {title} {Large spin accumulation signals in ultrafast magneto-optical experiments}} (\bibinfo {year} {2025}),\ \Eprint {https://arxiv.org/abs/2501.05838} {arXiv:2501.05838} \BibitemShut {NoStop}%
\end{thebibliography}
\end{document}